%% file: ms.tex
\documentclass[12pt,preprint]{aastex}

\usepackage{epsfig}
\usepackage{xspace}

\newcommand{\nmax}{n_{\max}}
\newcommand{\He}{ \mbox{He} }
\newcommand{\Hy}{ \mbox{H}  }
\newcommand{\sHe}{ \mathrm{He} }
\newcommand{\sHy}{ \mathrm{H}  }

\shorttitle{Paper 1}
\shortauthors{Bauman {\it et~al.}}

\defcitealias{F87}{F87}
\defcitealias{McAW78}{MW78}
\defcitealias{Drake93}{D93}
\defcitealias{HS98}{HS98}
\defcitealias{HB90}{HB90}
\defcitealias{Drake96}{D96}
\defcitealias{vanR}{R79}
\defcitealias{Peach}{P67}
\defcitealias{BS60a}{BS60a}
\defcitealias{BS60b}{BS60b}
\defcitealias{BSS99}{BSS99}
\defcitealias{Ferland98}{Fer98}
\defcitealias{BM38}{BM38}
\defcitealias{Pagel97}{P97}

\begin{document}


\title{$J$-resolved He~I emission predictions in the low-density limit}

\author{R. P.~Bauman, R. L.~Porter, G. J.~Ferland, and K. B.~MacAdam}
\affil{Department of Physics and Astronomy, University of Kentucky, Lexington, KY, 40506}
\email{rporter@pa.uky.edu}

\begin{abstract}
Determinations of the primordial helium abundance are used in precision 
cosmological tests. These require highly accurate He~I recombination rate 
coefficients. Here we reconsider the formation of He~I recombination lines 
in the low-density limit. This is the simplest case and it forms the basis 
for the more complex situation where collisions are important. The formation 
of a recombination line is a two-step process, beginning with  the 
capture of a continuum electron into a bound state and followed by radiative 
cascade to ground. The rate coefficient for capture from the continuum is 
obtained from photoionization cross sections and detailed balancing, while 
radiative transition probabilities determine the cascades. We have 
made every effort to use today's best atomic data. Radiative decay rates are 
from Drake's variational calculations, which include QED, fine structure, and 
singlet-triplet mixing. Certain high-$L$ fine-structure levels do not have a 
singlet-triplet distinction and the singlets and triplets are free to mix in 
dipole-allowed radiative decays. We use  
quantum defect or hydrogenic approximations to include levels higher than 
those treated in the variational calculations. 
Photoionization cross sections come from $R$-matrix 
calculations where possible. We use Seaton's method to extrapolate along 
sequences of transition probabilities to obtain threshold photoionization 
cross sections for some levels. For higher $n$ we use scaled hydrogenic 
theory or an extension of quantum defect theory. We create two 
independent numerical implementations to 
insure that the complex bookkeeping is correct. The two codes use different 
(reasonable) approximations to span the gap between lower levels, 
having  accurate data, and high levels, where scaled hydrogenic 
theory is appropriate. We also use different (reasonable) methods to account 
for recombinations above the highest levels individually 
considered. We compare these independent predictions to estimate the 
uncertainties introduced by the various approximations. Singlet-triplet
mixing has little effect on the observed spectrum. While intensities of
lines within multiplets change, the entire multiplet, the quantity normally
observed, does not. The lack of high-precision photoionization cross sections
at intermediate-$n$, low-$L$ introduces $\sim 0.5\%$ uncertainties in
intensities of some lines. The high-$n$ unmodeled levels introduce $\sim 1\%$
uncertainties for yrast lines, those having $L=n-1$ upper levels. 
This last uncertainty will not be present in actual nebulae since such high levels are
held in statistical equilibrium by collisional processes. We identify those
lines which are least affected by uncertainties in the atomic physics and
so should be used in precision helium abundance determinations.
\end{abstract}


\section{Introduction}
\label{sec:intro}
The spectra of hydrogen and helium emitted in the recombination process 
$A^{+} + e^- \to  A^\ast + \, \hbar\omega $ 
followed by subsequent cascades 
$A^* \to A'^*+ \hbar \omega'$, 
have long played a fundamental role in studies of cosmic chemical 
evolution. The relative intensities of the emission lines depend 
mainly on the abundances of $\Hy^+$ and $\He^+$, not on uncertain plasma 
conditions such as temperature and density, so ionic abundances can be 
determined with a precision that is limited instead by measurement errors and 
atomic theory. Much effort has gone into precision measurements of He/H 
abundance ratios with a particular emphasis on using the primordial 
abundance of He as a test of the Big Bang~\cite[hereafter~\citetalias{Pagel97}]{Pagel97}. 
This requires that theoretical emission spectra be understood to a precision better than $1\%$.

Calculation of the hydrogen recombination-cascade spectrum was one of the first applications of 
quantum mechanics to astrophysics~\cite[hereafter~\citetalias{BM38}]{BM38}. Hydrogen is a 
simple system, and it is thought that current predictions~\cite{StoreyHummer95}
are accurate to substantially better than $1\%$. The atomic physics of 
helium, being a two-electron system, is more complex. It was only much later 
that its recombination-cascade spectrum was first computed 
(see Brocklehurst 1972 for discussion), and recent studies have been published 
by  Smits~(1991,1996) and Benjamin~{\it et~al.}~(1999, hereafter~\citetalias{BSS99}). Each 
succeeding study improved the prior treatment of physical processes, mainly 
as the result of improved theoretical calculations of various rates. But the 
bookkeeping associated with solving the numerical problem involving several 
hundreds or thousands of levels is also intricate, and mistakes are almost 
unavoidable. Many of the successive papers found numerical errors in the 
preceding work.

This paper revisits the He~I recombination-cascade spectrum in the 
low-density limit. We make the following improvements. $\He^{0}$ has previously 
been modeled as distinct singlet and triplet systems with  $n\, {}^{2S + 1}\! L$ terms. The present
calculation utilizes fine-structure $n\, {}^{2S + 1}\! L_J$ levels. In $L \ge  3$ 
levels, however, the spin-orbit interaction leads to strong singlet-triplet 
mixing. We use Drake's (1996, hereafter~\citetalias{Drake96}) highly accurate calculations of 
the $J$-resolved transition probabilities, which take this singlet-triplet 
mixing into account. We carry out the calculation with $J$-resolved transitions 
twice: once with singlet-triplet mixing explicitly included ($ST$-mixing) and 
once with $LS$-coupling assumed ($LS$-coupling). 
Comparison of emission line intensities (or emission coefficients) allows us to ascertain directly the 
effects of including singlet-triplet mixing. Finally, to avoid  
bookkeeping errors, we do calculations with two independently developed codes to confirm predictions. 
The second code~\cite{PBFM} assumes pure $LS$-coupling and is not a $J$-resolved calculation. 
By summing the emissions from the $J$-resolved levels, we can compare the emission coefficients
to other multiplet-emission calculations. 

Based on the "principle of spectroscopic stability" (Condon \& Shortley 1991), only small
changes are to be expected in multiplet-average line intensities, either as a result of
allowance for $J$-splittings within $LS$-coupled terms or mixing between singlets and triplets.
This is because both of these effects can be expressed, at least to lowest order, in terms
of unitary transformations of the zero-order states, and the difference between the
sum-of-squares of electric-dipole matrix elements and the calculation of multiplet emission
or absorption strength hinges only on the tiny energy splittings involved. By the same token,
however, multiplet-average emission or absorption cannot be exactly independent of the
allowance for fine-structure and singlet-triplet mixing because of these very splittings,
and without explicit calculation the deviations, which are potentially important for accurate
interpretation astrophysical data, cannot be guessed.

Although extremely accurate atomic data now exist for the lower levels $\He^0$, we 
find that they do not extend to a high enough $n$ for the lower non-hydrogenic 
$L$ that are needed for definitive predictions of the spectrum. Various assumptions
are made to bridge the gaps between states with precise atomic data and those for
high $n$ and low $L$.  We identify the atomic data that introduce the greatest uncertainty  
in the final spectrum. Section~\ref{sec:atomicdata} discusses the necessary atomic 
physics and data sources. Section~\ref{sec:RRCP} describes the formation and solutions of 
the recombination-cascade problem. The results of this study are 
presented in Section~\ref{sec:RandU}, and conclusions are stated in Section~\ref{Conc}.  

\section{Atomic Data}
\label{sec:atomicdata}
The accuracy of the recombination and radiative cascade model presented here 
is determined mainly by the atomic data. A description of the relevant 
quantities, techniques, and references is given below. The high-precision 
calculations of~\citetalias{Drake96} are used extensively in the 
calculation of level energies, quantum defects, oscillator strengths, and 
matrix elements for $n \leq 10$. Extrapolations of the~\citetalias{Drake96} 
results are used in the calculation of some atomic data for the higher lying levels.

Here we are only interested in transitions between pairs of singly excited levels 
in helium sharing a $1s$ core configuration. For these levels the total orbital 
angular momentum $L$ equals the orbital angular momentum of the excited electron $\ell$. 
We will use the notation 
$\gamma_u \equiv \{n_u ,L_u ,S_u ,J_u \}$ 
for the initial (upper) level of an emission line and similarly 
$\gamma_l \equiv \{n_l ,L_l ,S_l ,J_l \}$ 
for the final (lower) level and 
$\gamma \equiv \{n,L,S,J\}$ 
for a level in general. 
We designate continuum levels with free electron energy 
$\varepsilon$ 
as
$\gamma (\varepsilon ) \equiv \{\varepsilon, L, S, J\}$.

\subsection{Level Energies}
We calculate the level energies in helium, depending on $n$ and $L$, by three methods. 
For levels $n \le 10$ and $L \le 7$, ionization energies $E$ are
obtained from the Rayleigh-Ritz variational relativistic calculations of~\citetalias{Drake96}. 
For all levels $L \ge 8$, the asymptotic multipole expansion method \cite{Drake93a,drach93} 
is used to calculate the (negative) eigenenergies $E_0$. 
Ionization energies are found from the relation $E = ( - E_0 - 4\, hc R_{\sHe^{2+}} )$
where $R_{\sHe^{2+}}$ is the Rydberg constant for an electron-plus-alpha-particle system.
For levels $n \ge 10$ and $L \le 7$, ionization energies are found from the Ritz quantum defect 
expansion~\citepalias{Drake96}. These energies include all relativistic and quantum electrodynamic~(QED) 
corrections to the nonrelativistic eigenenergies through order $\alpha^4_{\mathrm{fs}}$, 
where $\alpha_{\mathrm{fs}}$ is the fine-structure constant.
Overlap at the boundaries of the three $nL$ regions allows us to verify 
the accuracy of our implementation.

For each $n$ and $L$, the energies of the two levels with 
$J=L$~({\it e.g.} $n\, {}^3\! L_L$ and $n\, {}^1\! L_L$) 
are shifted by the off-diagonal fine-structure ($J$-resolved) matrix 
elements connecting these two levels~\cite[hereafter~\citetalias{McAW78}]{McAW78} to give 
the singlet-triplet mixing energies. Quantum defects $\delta$ 
and effective quantum numbers $\nu = n - \delta$ are then calculated 
from the modified level energies.

Exact analytical solutions to the nonrelativistic Schr\"{o}dinger equation are 
known for two-body systems ($e.g.$ atomic hydrogen). For helium, approximate 
solutions based on the Rayleigh-Ritz variational principle are 
now available~\citepalias{Drake96} and are essentially exact.
Relativistic and QED  corrections are then added, including both
diagonal and off-diagonal matrix elements of spin-orbit and spin-other-orbit interactions
(\citetalias{Drake96},~Drake~1993b).
It is these off-diagonal matrix elements that mix levels of different
total spin $S$ and are responsible for the breakdown of $LS$-coupling.

For all levels with $L \ge 8$, the asymptotic expansion 
describes the interaction of the Rydberg electron with the $\He^+$ core in terms of 
core-polarization multipole moments \cite{drach93}.
This approximation agrees with the full variational calculation at $L=7$ and further improves with 
increasing $L$.

The ionization energies of excited 
helium Rydberg levels deviate from hydrogenic values and 
may be represented by
\begin{equation}
E(\gamma ) = \frac{hc R_{\sHe^+}}{(n - \delta (\nu))^2} \,  
\end{equation}
\noindent
$R_{\sHe^+}$ is the Rydberg constant for 
the reduced mass of the electron-$\He^+$ system. The quantum 
defects $\delta(\nu)$, in addition to having a dependence on $S$,~$L$, and $J$, also
depend weakly on $\nu$. $\nu$ is found by an iterative 
solution to the equation, $\nu = n - \delta (\nu)$, where in the Ritz expansion 
(Edl\'{e}n~1964)
\begin{equation}
\delta (\nu ) = \delta_0 + \frac{\delta_2 }{(n - \delta (\nu))^2} + 
\frac{\delta_4 }{(n - \delta (\nu))^4} + \cdots 
\end{equation}
The constant coefficients $\delta_i$ used here are given by~\citetalias{Drake96}.

\subsection{Bound-bound transitions}
The emission oscillator strength $f_{ul}$ (dimensionless) and the 
spontaneous radiative transition rate coefficient (Einstein $A$; s$^{-1}$) 
are principal atomic quantities related to line strengths for 
transitions between an initial upper level $\gamma_u$ and a final lower 
level $\gamma_l$. The Einstein $A$ coefficients are the most convenient quantity for 
calculating the elements of the cascade matrix while theoretical atomic work 
usually refers to oscillator strengths. The relationship between the two for the 
electric dipole transitions in SI units is:
\begin{equation}
A_{ul} = \frac{2\pi e^2}{m_e c\varepsilon_o \lambda^2} f_{ul},
\end{equation}
\noindent
where $\lambda$ is the vacuum wavelength.

\subsubsection{Drake's emission oscillator strengths}
\label{DrakeOscStr}
For transitions with 
$\Delta S = 0$, 
$n_l \le n_u \le 10$, 
and both 
$L_u$ and $L_l \le 7$, 
including those with 
$\Delta n = 0$,  
the tabulated emission oscillator strengths of~\citetalias{Drake96} are used.  
These are high precision 
$J$-resolved calculated values which include QED, relativistic fine-structure and 
non-fine-structure corrections. The largest relativistic correction comes 
from singlet-triplet mixing between levels with the same $n$,~$L$, and $J$. 
In addition, ~\citetalias{Drake96} provides oscillator strengths and Einstein $A$ 
coefficients both assuming pure $LS$-coupling (\textit{i.e.} no singlet-triplet mixing) and 
with singlet-triplet mixing included. Emission oscillator strengths for 
transitions with $f_{ul} \le 10^{ - 6 }$ are omitted, but we calculate them by 
a Coulomb approximation method described later.

\subsubsection{Extrapolated emission oscillator strengths}
\label{sec:extrapolation}
For transitions with 
$\Delta S = 0$, and 
$n_u \ge 11$ and 
$n_l \le 7$ and either 
$L_u \le 6$ or 
$L_l \le 6$, 
the emission oscillator strengths are derived by extrapolating those of~\citetalias{Drake96}. 
To find the emission oscillator strength $f_{ul}$ we extrapolate the series $f_{jl}$ with 
$\gamma_j \equiv \{n_j ,S_u ,L_u ,J_u \}$ for
$n_j = n_l + 1,n_l + 2, \cdots ,10$. 
This series is fitted as 
$\ln (\nu_j^3 \, f_{jl} ) = a + bx + cx^2$, with 
$x = \ln (E_l / E_{jl})$, 
as suggested by Hummer~{\&}~Storey (1998, hereafter~\citetalias{HS98}).
The oscillator strength dependency for large $n$, $f\sim \nu^{ - 3 }$, is represented by the $\nu_j^3$ factor. 
Parameters $a, b, $ and $c$ are determined by the fit. Here $E_l$ is the ionization energy
of level $\gamma_l$ and $E_{jl}$ is the energy difference between levels
$\gamma_j$ and $\gamma_l$. For some series with small $n_j$, the lowest lying members are
omitted from the fit to obtain a better estimate of the parameters.

\subsubsection{Coulomb approximation method}
\label{CoulombOscStr}
A Coulomb approximation method~\cite[hereafter~\citetalias{vanR}]{vanR}
is used to calculate the oscillator strengths for all remaining transitions except 
for those with $L_u = n_u -1 $ or $L_l = n_l - 1$. In transitions involving ${}^{1}P$ 
levels, the method is extended to account for negative quantum defects, 
a special case not addressed in~\citetalias{vanR}. Emission oscillator strengths for weak 
transitions not included in~\citetalias{Drake96} are calculated using this method. This 
simple method is particularly suitable for transitions involving high Rydberg 
levels with 
$\nu_l , \nu_u > 20$ and 
$\Delta \nu \ll \nu_l , \nu_u $ where 
$\Delta \nu = \nu_u - \nu_l$. 
It agrees with the Bates \& Damgaard~(1949) 
results for $\nu_l,\nu_u < 20$ and with hydrogenic results for which $\nu$ takes 
an integer value. The method is based on the observation that, 
for fixed values of $\Delta \nu$, $L_{u}$, and $L_{l }$,
the variation
of the radial integrals $R_{\gamma_l }^{\gamma_u }$
with $\nu_{u}$ (or $\nu_{l}$) is very slow.
Therefore, one of the principal quantum numbers may be taken to be an 
integer, and the results may be obtained accurately by interpolation.

\subsubsection{Hydrogenic oscillator strengths}
\label{sec:Hoang-Bing}
\label{HydrogenicXSec}
The remaining oscillator strengths are all taken to be hydrogenic. The 
emission oscillator strengths are hydrogenic if quantum defects of the upper and lower levels 
are nearly zero. The radial integrals $R_{\gamma_l }^{\gamma_u}$ necessary to find 
the oscillator strengths for these transitions are calculated by 
the hydrogenic solution of Hoang-Bing~(1990, hereafter \citetalias{HB90}), which is an accurate and 
efficient method to calculate the exact analytical solution of Gordon~(1929).

\subsubsection{$J$-resolved oscillator strengths}
\label{J-resolved}
The methods of~\citetalias{vanR} or \citetalias{HB90} provide radial integrals and are 
used to calculate the $J$-resolved emission oscillator strengths. 
The (mean) oscillator strength is defined by
\begin{equation}
f_{ u l } = \delta_{S_u S_l } \frac{2\mu \omega }{3\hbar}
\sum\limits_{M_l = - J_l }^{J_l} 
{\frac{1}{2J_u + 1}
\sum\limits_{M_u = - J_u }^{J_u } 
{\vert \langle } } 
\,\gamma_u \;M_u \vert  \mathord{\buildrel{\lower3pt\hbox{$\scriptscriptstyle\rightharpoonup$}}\over {r}} \vert \gamma_l \;M_l 
\rangle \vert^2.
\end{equation}
\noindent
Here $\omega = (E_u - E_l ) / \hbar$ is the transition frequency, $\mu$ is 
the reduced mass, and $\delta $ is the Kronecker delta. When the angular 
momentum operators $L$ and $S$ that sum to $J$ are decoupled,
the oscillator strength may be written 
\begin{equation}
f_{ u l } = \delta_{S_u S_l } 
\frac{2\mu \omega }{3\hbar }\frac{(2J_l + 1)\,L_ > }{(2L_u + 1)}
\left\{\matrix{ {L_u}& {1}& {L_l} \cr {J_l}& {S_u}& {J_u} \cr}\right\}^2
\, \left( {\int\limits_0^\infty {dr \; \phi^\ast (\gamma_u ;r) \; r \; \phi (\gamma_l ;r)} } \right)^2.
\end{equation}
\noindent
Here $L_>=\mathrm{max}(L_u,L_l)$ and the {\{}{\}} factor is a 
Wigner 6$j$ symbol~(see Edmunds~1960). The expression in parentheses is the radial integral  $R_{\gamma_l }^{\gamma_u}$
discussed earlier. The function $\phi ( \gamma_l ;r )$ 
is the radial part of the wavefunction
$\Psi (\gamma ;\vec{r})= r^{-1}\, \phi (\gamma ;r) \, Y^{M_L}_L(\Omega)$.
Oscillator strengths for $\Delta S \ne 0$ (allowed by singlet-triplet mixing) 
are discussed in the following subsection.

\subsubsection{Oscillator strengths under singlet-triplet mixing}
The largest relativistic correction to helium oscillator strengths comes from 
singlet-triplet mixing. This occurs most significantly between the
two nominally singlet and triplet $LS$-coupled components 
with $J = L$  of a given $nL$ ({\it e.g.} $4\, {}^3D_2$ and $4\, {}^1D_2$). The largest 
component to the correction is due to the magnetic inner-spin 
outer-orbit interaction. The $P$ and $D$ series are only 
very weakly mixed, because the singlet-triplet basis states are widely 
separated by the electron exchange interaction. Substantial mixing occurs in $F$ levels,
where exchange is much weaker, and for $L \ge 4$ 
the two $J = L$ energy eigenstates in each $nL$ multiplet are almost equal 
mixtures of singlet and triplet character. Oscillator strengths are obtained 
from the rediagonalization of the $(2 \times 2)$ matrices for these pairs of levels 
as described by the mixing angle $\theta$~\citepalias{Drake96}. The mixed-spin 
wave functions $\Psi$ obtained by rediagonalization from the unmixed 
wavefunctions $\Psi_0$ are
\begin{equation}
\begin{array}{l}
 \Psi (n\, {}^1L_L ) = + \Psi_0 (n\, {}^1L_L )\cos \theta + \Psi_0 (n\, {}^3L_L )\sin \theta   \\ 
 \Psi (n\, {}^3L_L ) = - \Psi_0 (n\, {}^1L_L )\sin \theta + \Psi_0 (n\, {}^3L_L )\cos \theta . \\ 
 \end{array}
\end{equation}
\noindent
We retain the traditional notation for the mixed-spin wavefunctions
with the understanding that only in the limit $\theta \rightarrow 0$ are the 
indicated multiplicities exact.
The corresponding corrected (singlet-triplet mixed) oscillator strengths 
$\tilde f_{\gamma {\gamma }'}$ for the singlet ($s$) and triplet ($t$) 
components of a $\gamma \to {\gamma }'$ transition are written in terms of 
the unmixed oscillator strengths $f_{\gamma {\gamma }'}$ as
\begin{equation}
{
\begin{array}{*{20}c}
\tilde {f}_{\gamma {\gamma }'}^{ss} = \omega_{\gamma {\gamma }'}^{ss} 
(X_{\gamma {\gamma }'}^{ss} \cos \theta_\gamma \cos \theta_{\gamma }' + 
X_{\gamma {\gamma }'}^{tt} \sin \theta_\gamma \sin \theta_{\gamma }' )^2 \hfill \\

\tilde {f}_{\gamma {\gamma }'}^{tt} = \omega_{\gamma {\gamma }'}^{tt} 
(X_{\gamma {\gamma }'}^{ss} \sin \theta_\gamma \sin \theta_{\gamma }' + 
X_{\gamma {\gamma }'}^{tt} \cos \theta_\gamma \cos \theta_{\gamma }' )^2 \hfill \\

\tilde {f}_{\gamma {\gamma }'}^{st} = \omega_{\gamma {\gamma }'}^{st} 
(X_{\gamma {\gamma }'}^{ss} \cos \theta_\gamma \sin \theta_{\gamma }' - 
X_{\gamma {\gamma }'}^{tt} \sin \theta_\gamma \cos \theta_{\gamma }' )^2 \hfill \\

\tilde {f}_{\gamma {\gamma }'}^{ts} = \omega_{\gamma {\gamma }'}^{ts} 
(X_{\gamma {\gamma }'}^{ss} \sin \theta_\gamma \cos \theta_{\gamma }' - 
X_{\gamma {\gamma }'}^{tt} \cos \theta_\gamma \sin \theta_{\gamma }' )^2 \hfill \\
\end{array}
}
\end{equation}
\noindent 
where 
$X_{\gamma {\gamma }'}^{ss} = (f_{\gamma {\gamma }'}^{ss} / \omega_{\gamma {\gamma }'}^{ss} )^{1 / 2}$, 
and similarly for 
$X_{\gamma {\gamma }'}^{tt}$, 
and 
$\omega_{\gamma {\gamma }'}^{ss}$, $\omega_{\gamma {\gamma }'}^{tt}$, $\omega_{\gamma {\gamma }'}^{st}$, 
and 
$\omega_{\gamma {\gamma }'}^{ts}$ 
are the (modified) transition frequencies.

For low lying levels with $n \le 10$ and $L \le 9$ we use tabulated 
mixing angle data~\cite{Drake96}. Higher lying levels with $n \ge 11$ and $L \ge 7$ are 
nearly equally mixed and we use $\theta = 45^{\circ}$. For levels with $n \ge 11$ 
and $L > 3$ , the mixing angle is approximately constant for increasing $n$ in
each $L$ series and we use the $n=10$ value of the mixing angle for all higher 
levels. For levels $n \ge 11$ and $L \le 3$, mixing angles are slowly 
monotonically decreasing with increasing $n$. For these levels we solve the secular determinant 
for the fine-structure splitting in a configuration $1sn\ell$ with the exchange 
integral included along the diagonal~\citepalias{McAW78}. These agree quite well with a simple 
extrapolation of the lower-level mixing angles in each of the $nL$ series. The 
pure $LS$-coupling calculation is equivalent to making the assignment $\theta = 0$.

\subsubsection{Included non-dipole transitions and oscillator strengths}
Several non-dipole-allowed $n = 2 \to 1$ and $n = 2 \to 2$ transitions are 
included to facilitate comparison with previous works. Einstein $A$ coefficients for the 
non-dipole transitions are from the literature as follows: the two photon transition  
$2\,{}^{1}S_{0} \to 1\,{}^{1}S_{0}$ is from Drake (1979); 
$2\,{}^{3}S_{1} \to 1\,{}^{1}S_{0}$ is from Hata~\&~Grant~(1981);
$2\,{}^{3}P_{1} \to 1\,{}^{1}S_{0}$ and 
$2\,{}^{3}P_{2} \to 1\,{}^{1}S_{0}$ are from Lin~{\it et~al.}~(1977) ;
$2\,{}^{3}P_{0} \to 1\,{}^{1}S_{0}$ is from Drake~(1969). 
The remaining oscillator strengths are from~\citetalias{Drake96}.

\subsection{Radiative Recombination Rates}
\label{sec:rrr}
Radiative recombination rates are obtained from the He~I photoionization 
cross sections by the method of detailed balancing~(Seaton~1959). The number 
of recombinations to a level $\gamma$ per unit volume per unit time is given by 
$\alpha (T;\gamma )\, n_e n_{\sHe^+}$,
where 
$n_e$ and $n_{\sHe^+}$ 
are the electron and helium-ion number densities, respectively. The radiative 
recombination coefficients 
$\alpha (T;\gamma )$
for the process~$\He^+ + e^- \to \He(\gamma )+\hbar \omega$~are 
given by the Milne relation~(see Osterbrook~1989), appendix 1)
\begin{equation}
\alpha (\gamma ;T) = \frac{c\alpha_{\mathrm{fs}}^3 }{\sqrt \pi }\,\frac{(2L + 1)(2S + 1)}{4}\,
\beta^{3 / 2}\nu^{ - 4}\,
\int\limits_0^\infty {d\varepsilon\,(1 + \nu^2\varepsilon )^2\,e^{ - \beta \varepsilon }} \,\sigma (\gamma ;\varepsilon )
\end{equation}
\noindent
where 
$\sigma (\gamma ;\varepsilon )$ 
is the photoionization cross section 
from level    
$\gamma $ 
yielding a free electron having energy    
$\varepsilon$ 
(in Rydberg units $hc R_{\sHe^+}$),
$\beta = hc R_{\sHe^+} / k_B T$ 
for temperature $T$ and Boltzmann constant $k_{B}$. 
The Maxwell-Boltzmann distribution function is represented by 
$(1 + \nu^2\varepsilon)^2\,e^{ - \beta \varepsilon}$. 
The integration scheme used for detailed balancing is outlined by 
Burgess~(1965) and Brocklehurst~(1972). For dipole transitions 
$\sigma (\gamma ;\varepsilon)$ 
is the sum of two partial photoionization cross sections to the two dipole-allowed $\Delta L = \pm 1$ continua:
$\sigma ( \gamma ; \varepsilon ) = \sigma_{\mathrm{p}} (\gamma ;\varepsilon, L + 1) + \sigma_{\mathrm{p}} (\gamma ; \varepsilon, L - 1 )$. 
If $L=0$, the second term is omitted.

Radiative recombination rates are the most uncertain quantities in the model 
calculation. For the lowest lying levels with $n \le 7$ and $L=0$ or $1$ the cross 
sections of Fernley~\textit{et~al.}~(1987, hereafter~\citetalias{F87}) are used. Certain photoionization cross sections are 
missing from that work, and for these, as well as for levels with $n \le 9$ and $L \le 2$, 
the cross sections of Peach~(1967, hereafter~\citetalias{Peach}) are used.

\subsubsection{Hummer \& Storey Recombination Rates for $n \ge 25$}
\label{HS25}
\begin{figure*}[htbp]
\resizebox*{\textwidth}{!}{\includegraphics*[bb = 00 96  612 712]{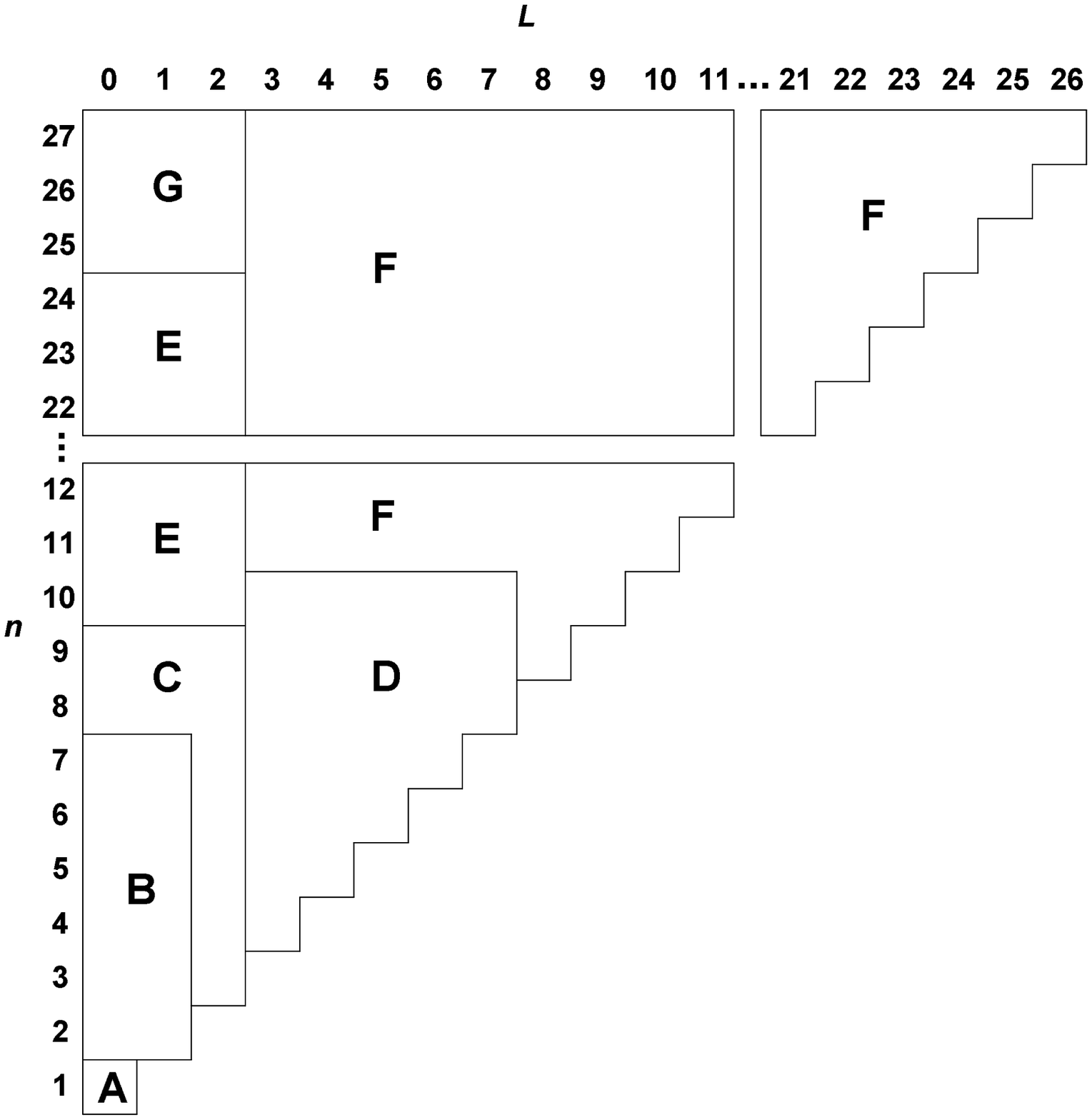}}
\caption[PhotoIonization Methodes]
{
Graphical representation of the methods used for photoionization cross sections.
The letters represent methods as follows:
A)~TOPbase;
B)~renormalized TOPbase (Section~\ref{RenormalizedXSec}); 
C)~renormalized Peach (Section~\ref{RenormalizedXSec});
D)~renormalized hydrogenic (Section~\ref{RenormalizedXSec});
E)~rescaled hydrogenic (Section~\ref{RescaledHydroXSec}); 
and
F)~pure hydrogenic (Section~\ref{HydrogenicXSec});
G)~rescaled hydrogenic (Section~\ref{HS25});
}
\label{fig1}
\end{figure*}

For the higher lying levels 
$n \le 24, L \le 2$, 
hydrogenic recombination rates (Burgess \& Seaton~1960a,~1960b,
hereafter~\citetalias{BS60a} and~\citetalias{BS60b})
are calculated and then scaled by the 
ratio of helium and hydrogen threshold photoionization cross sections. 
For levels with 
$n \ge 25$ and $L \le 2$, 
hydrogenic recombination rates are used with scale factors given by~\citetalias{HS98}. 
For levels  
$n \le 10$ and $L \ge 3$, 
or for all levels
$L \ge 7$, 
pure hydrogenic recombination rates are used. 
Hydrogenic rates for $L \ge 4$ agree with those of helium to at least three figures (\citetalias{HS98}). 
The methods used to calculate the radiative recombination rates for individual $nL$ levels are depicted in Figure~\ref{fig1}.

\subsubsection{Photoionization cross sections}
The photoionization cross section for photons of arbitrary polarization in 
terms of the differential oscillator strength is given by 
(see, for example, Friedrich~1990)
\begin{equation}
\sigma (\varepsilon ) = 4\pi^2 a_0^2\, \alpha_{\mathrm{fs}} \, \frac{d f}{d\varepsilon } 
= (4.033643\times 10^{-18} \mathrm{~cm}^2 ) \; \frac{d f}{d\varepsilon } \;
\end{equation}
\noindent
where $a_0$ is the Bohr radius.
For photoionization from an initial (lower) bound state with 
$n_{l}, \ell_{l}$ 
to a final (upper) continuum state with angular momentum 
$\ell_u$,
the non-$J$-resolved (mean) photoionization 
differential oscillator strength is  
\begin{equation}
\frac{df_{n_l \ell_l ,\varepsilon \ell_u } }{d\varepsilon } = \frac{2\mu }{3 \hbar}\omega \,\frac{\ell_>}{2\ell_l + 1}
\left( {\int\limits_0^\infty {dr\;\phi^\ast_{n_l \ell_l } (r)\;r\;\Phi_{\varepsilon \ell_u } (r)} } \right)^2.
\end{equation}
\noindent
Here, the initial bound state radial wavefunction is 
$\phi_{n_l \ell_l}(r)$
and the  final (energy normalized) continuum-state radial wavefunction is  
$\Phi_{\varepsilon \ell_u }(r)$.

\subsubsection{TOPbase photoionization cross sections}
For levels $n=2$ to $7$ and $L=0$ or $1$, the photoionization cross sections 
used for the calculation of the recombination rates are obtained from the 
Opacity Project~(\citetalias{F87}) as deposited in the database 
TOPbase\footnote[1]{http://vizier.u-strasbg.fr/topbase/topbase.html}
(Cunto~\textit{et~al.}~1993).
These are labeled with B in Figure~\ref{fig1}.
The photoionization cross sections of~\citetalias{F87} are \textit{ab initio} 
close-coupling calculations using the $R$-matrix 
method~(Berrington~\textit{et~al.}~1974,~1978,~1987)
of the scattering of an electron from a helium ion. 
For those photoionization cross sections 
missing from the database we use the method of~\citetalias{Peach}.

\subsubsection{Peach photoionization cross sections}
For levels $n = 3$ to $9$ and $L = 0$ or $2$, the partial photoionization cross sections 
are obtained from~\citetalias{Peach}. 
These levels are labeled with C in Figure~\ref{fig1}.
The method of~\citetalias{Peach} is based on the quantum defect representation of Coulomb wavefunctions
and boundary conditions of~\citetalias{BS60a}. 
It is applicable for states with the initial 
bound-state principal quantum number $n_i \le 12$ and may be 
used to calculate partial photoionization cross sections with initial orbital 
quantum number $\ell = L \leq 2$. 
These partial photoionization cross sections are sufficient to calculate the 
recombination coefficients for $S$, $P$,~and $D$ states. 
A functional form  $\nu (E)$ is first found from the 
quantum defects for each series to calculate the required first derivative of $\nu$ and 
the non-hydrogenic part of the continuum phase beyond the photoionization threshold.

The form of the partial photoionization cross section 
$\sigma_{\mathrm{p}} (\nu, \ell ;\varepsilon, \ell \pm 1  ) $ 
is given by 
\begin{equation}
\sigma_{\mathrm{p}}(\nu, \ell ;\varepsilon, \ell \pm 1  ) = 
\frac{8 {\kern 1pt} \alpha_{\mathrm{fs}} a_0^2  \nu^3}
{3 \zeta (\nu, \ell)}(1 + \nu^2 \varepsilon )^{ - 3} \, C^{\ell}_{\ell \pm 1} 
\left[ 
   { G(\nu, \ell; \varepsilon,  \ell \pm 1)
   \cos \,\pi (\nu + \mu' (\varepsilon ) + 
   \chi (\nu, \ell ; \varepsilon,  \ell \pm 1))} 
\right]^2.
\end{equation}
\noindent
Here 
$ \mu' (\varepsilon )$
is the continuum-state quantum defect phase and 
$C^\ell_{\ell \pm 1} = \ell_> / (2\ell + 1)$  
are coefficients (\citetalias{BS60a}) obtained from the 
integrations over spin and angular co-ordinates.
~\citetalias{Peach} tabulates the necessary amplitudes 
$ G (\nu, \ell ; \varepsilon,  \ell \pm 1)$ 
and
$\zeta (\nu, \ell)$
and the non-hydrogenic phase
$ \chi (\nu, \ell ; \varepsilon, \ell \pm 1 )$ 
for ejected-electron energies 
$\varepsilon \le 2 hc R_{\sHe^+}$. 
At the temperatures considered here, 
by far the largest contribution to the recombination rates
is from the first few eV, 
so that ion-core excitations and two-electron processes 
do not contribute to the integral.

\subsubsection{Hydrogenic photoionization cross sections}
\label{sec:hydro-cs}
For levels in which $n$ and $L$ are large enough, 
the core electron fully screens the nucleus, and
exact analytic hydrogenic cross sections 
are used to calculate recombination rates. 
These levels are labeled with D, E, F, and G in Figure~\ref{fig1}.
Cross sections for this process are given by \citetalias{BS60a}, 
and the implementation described by \cite{Brocklehurst72} is used.

\subsubsection{Hydrogenic cross sections}
\label{RescaledHydroXSec} 
For  $n > 10$ and $L < 4$, we use scaled hydrogenic cross 
sections. The scale factor is an extrapolation as $n\rightarrow \infty$
of the ratio $\alpha_{\sHe}(\gamma)/\alpha_{\sHy}(nL)$, 
where $\alpha_{\sHe}(\gamma)$ and $\alpha_{\sHy}(nL)$  are the
helium and hydrogen recombination coefficients, respectively. 
We fit these series of ratios 
$\alpha_{\sHe}(\gamma)$/$\alpha_{\sHy}(nL)$ by
\begin{equation}
\frac{ \alpha_{\sHy} } {\alpha_{\sHe} } = a + \frac{b}{n^2} + \frac{c}{n^4}
\end{equation}
\noindent
where the third term is only used for the ${}^3P$ series.
Our results for $\alpha_{\sHe}$/$\alpha_{\sHy}$ agree well 
with HS98 at $n = 25$ for the singlet and triplet $S$,~$P$, and $F$ 
series but disagree for the singlet and triplet $D$ series by about 2.0\%.

\subsubsection{Renormalizing photoionization cross sections}
\label{RenormalizedXSec}
\citetalias{HS98} concludes that neither the photoionization cross sections from Peach's 
Coulomb method nor those of the Opacity Project are ideal.
Extrapolation of the {\it absorption} oscillator strengths of~\citetalias{Drake96},
based on Seaton's Theorem~\cite{Seaton58} and as discussed in Section~\ref{sec:extrapolation},
to $E_{jl} = E_l$ yields the photoionization cross sections at threshold ($\varepsilon = 0$).
These differ, for $L\le 3$, by up to $5.0\%$ from those of~\citetalias{Peach} and the Opacity Project.
We use the extrapolated threshold values to renormalize the continuum cross sections.
Similarly renormalized hydrogenic cross sections are used for levels $n \le 10$ and $L \ge 3$.

\subsubsection{$J$-resolved photoionization cross sections}
The \citetalias{Peach}, TOPbase and hydrogenic photoionization cross sections are not 
$J$-resolved. The analysis used to find $J$-resolved in terms 
of non-$J$-resolved photoionization cross sections is similar to the above 
analysis of  oscillator strengths.
The (mean) partial photoionization cross section is given by
\begin{equation}
\sigma_{\mathrm{p}}( \gamma_l, \gamma_u;  \omega )
= \frac{ 2 \mu \omega }{ 3 \hbar }
\sum\limits_{M_u = - J_u}^{J_u} 
\frac{1}{2J_l + 1}
\sum\limits_{M_l = - J_l}^{J_l} 
\vert \langle \, 
\phi_{\gamma_l M_l} \vert \vec{r} \vert \Phi_{{\gamma_u }{M_u}} 
\rangle \vert^2.
\end{equation}
\noindent


\noindent When the bound-free radial integrals can be explicitly calculated, the $J$-resolved (mean) total
photoionization cross section may be written as

\begin{equation}
\sigma ( \gamma; \omega ) = \sum\limits_{{L_u} = L_l\pm 1} \frac{2\mu \omega }
{3\hbar }\frac{L_ > }{(2L_l + 1)}
\,
\sum\limits_{J_u}
(2{J_u} + 1)
\left\{\matrix{ {L_l}& {1}& {L_u} \cr {J_u}& {S}& {J_l} \cr}\right\}^2
\,
\left( {\int\limits_0^\infty {dr\;\phi_{\gamma_l}^\ast (r)\;r\;\Phi_{{\gamma_u }} (r)} } \right)^2.
\label{this_equation}
\end{equation}

\noindent Equation \ref{this_equation} cannot, however, be used to calculate $J$-resolved cross sections
from pre-calculated non-$J$-resolved cross sections, such as those from TOPbase.  In this case, we produce 
$J$-resolved cross sections by apportioning the non-$J$-resolved cross sections according to the statistical weight
of the states within the lower term, as follows:

\begin{equation}
\sigma ( \gamma; \omega ) = \frac{2J + 1}{(2L + 1)(2S + 1)} \sigma_{TOP} ( nLS; \omega).
\end{equation}

\subsubsection{Recombination to levels with $n$ greater than $\nmax$ }
\label{topoff}
In the low density limit, an infinite number of levels $k$ must be considered.
The largest principal quantum number $n$ for explicitly considered levels is $\nmax$.
Simple truncation of the system at $\nmax$, however, would fail 
to account for the recombinations to and cascades from all higher levels, causing
an underestimation of emission coefficients. The recombination remainder $\alpha_{\mathrm{rem}}$,
the sum of the convergent infinite series of recombination to higher levels,
must therefore be artificially added to the direct recombination of the explicitly
treated levels. The recombination remainder is calculated by using an
approximation method described by Seaton (1959).

While recombination coefficients into a given $n$ are largest for low to moderate 
angular momenta and then sharply decline for greater angular momenta, 
\textit{effective} recombination into a given $n$---the sum of direct recombination and cascades
from higher levels---will be distributed among $\ell$ very nearly 
according to statistical weight $2\ell+1$.  In our treatment, we apportion 
$\alpha_{\mathrm{rem}}$ according to the statistical weights of the
separate $\ell$ levels with 
$ n = \nmax $ 
and add it to the \textit{direct} recombination  
$\alpha(\nmax,\ell)$ 
of the respective levels, so that the resultant recombination rate is given by 
\[
\alpha (\nmax , \ell) \rightarrow \alpha(\nmax ,\ell) + \left( {\frac{2\ell + 1}{\nmax^2 }} \right) \alpha_{\mathrm{rem}}.
\].

The second term in the above sum, which we refer to here as ``topoff'', is large
compared with the direct recombination (first term), and the difference is 
greatest for high $\ell$ levels. (Levels having $\ell=n-1$ are called ``yrast'' levels; see Grover 1967.)
In the low-density limit, an uncertainty is introduced by the addition of topoff, because
the levels are not actually statistically distributed.  This uncertainty
is minimized by employing the largest possible $\nmax$.

\section{Radiative Recombination Cascade Problem}
\label{sec:RRCP}

\subsection{Case~A and Case~B}
Baker \& Menzel~(1938) proposed two limiting cases of Lyman line optical depth in the interstellar medium (ISM).
The Case~A approximation assumes that the line-emitting region is optically thin and that
radiative excitation from the ground state is unimportant. The Case~B approximation
assumes that Lyman line photons originating from $n>2$ scatter often enough that they are degraded
to Balmer lines and Ly$\alpha$. Baker \& Menzel found that Case~B more closely reproduced
observations of hydrogen emission from the ISM than did Case~A.  In helium, singlet levels have the same
Case~A - Case~B distinction, but triplet levels, having no resonance lines, do not.  The
present calculation considers only the Case~B approximation.

\subsection{Rate-equation formalism}
In the steady-state, low-density, zero-incident-radiation limit we have the 
following balance equations for levels $k$ of $\sHe^0$: 
\begin{equation}
n_e n_{\sHe^+} \alpha(k;T) = \sum\limits_{E_l < E_k } {n_k A_{kl} } - 
\sum\limits_{E_j > E_k } {n_j A_{jk} } 
\end{equation}
\noindent
where $A_{pq}$ is the transition probability (s$^{ - 1})$ from level $p$ to 
level $q$, $n_{e}$ and $n_{\sHe^+}$ are the local electron and singly ionized 
helium number densities (cm$^{ - 3})$, $n_{k}$ is the number density of helium
atoms in level $k$ (cm$^{ - 3})$, and $\alpha(k;T)$ is the recombination 
coefficient (cm$^{3}$~s$^{ - 1})$ to level $k$ at temperature $T$(K).

The set of $k_{\max}$ balance equations (where 
$k_{\max}$ is the number of levels considered in the calculation)
can be solved for the vector of level densities 
$(n_1 , n_2 ,  n_3 , \ldots , n_k , \ldots , n_{k_{\max}} )$. 
With the level densities known, local line {\it emission coefficients} $4\pi j_{\lambda}/n_{e}n_{\sHe^+}$ 
for the radiation at wavelength 
$\lambda = hc/E_{kl}$, 
where 
$E_{kl}=E_{k }-E_{l }$, 
are
\begin{equation}
\frac{4\pi j_{\lambda}}{n_{e}n_{\sHe^+}} = \frac{n_k}{n_{e}n_{\sHe^+}} A_{kl} E_{kl} 
\end{equation}
\noindent
where $j_{\lambda}$ are the corresponding emissivities (erg~cm$^{-3}$~s$^{-1}$).
The emission coefficient is conventionally given in units of 
erg~cm$^{3}$~s$^{-1}$. 
(The conversion to SI units is 
1~erg~cm$^{3}$~s$^{-1} = 10^{-13}$~J~m$^{3}$~s$^{-1}$.)
The total intensity of the line (erg~cm$^{ - 2}$~s$^{ - 1}$) is the local emissivity
integrated over the depth of the line-emitting region.

\section{Results and uncertainties}
\label{sec:RandU}
We discuss our results for a single prototype 
case with a temperature of $10,000$~K and with particle densities  
$n_{e}\, = 1\mathrm{~cm}^{ - 3 }$ and $n_{\sHe^+}\, = 1 \mathrm{~cm}^{ - 3 }$.
Collisional interactions are ignored in this low-density limit.

\subsection{Absence of singlet-triplet mixing effects in multiplets}
A comparison of the emission coefficients, $4\pi j_{\lambda}/n_{e}n_{\sHe^+}$,
of the components of representative multiplets for singlet-triplet mixing
and for pure $LS$-coupling is presented in Table~\ref{table:J-Results}: there are some differences.
For transitions having $L_l = 0$ or $1$, which encompasses all of the ultraviolet 
and most of the strongest visible and longer wavelength lines, the differences 
in the emission coefficients are negligibly small. 
Many of the emission coefficients of longer wavelength lines ($L_l \ge 2$) 
show a strong sensitivity to the presence of singlet-triplet mixing.
Of course, intercombination lines ($\Delta S \ne 0$) are also then present. 
Large changes in emission coefficients when 
singlet-triplet mixing is included are almost entirely due to 
branching ratios as opposed to occupation numbers.
Further, any small differences in the 
occupation numbers do not ``accumulate'' along cascade paths 
and affect subsequent emissions.

The Doppler widths at temperatures of order $10,000$~K, typical in the ISM,
are such that, for most of the strongest IR and visible 
lines, the individual $J$ components are not resolvable.
Thus, Table~\ref{table:J-Results} also gives the summed multiplet emission coefficients.
These are not significantly affected by including singlet-triplet mixing. 
Therefore, in the remaining sections we will use pure $LS$-coupling
and provide multiplet emission coefficients.

\subsection{Effects of topoff and $\nmax$ on convergence}
The full problem with an infinite number of levels cannot be solved exactly.
There are two aspects of the effect of truncation---the modeling of a 
finite number of levels---on our results: One is the choice of $\nmax$, the 
highest principal quantum number used. The other (topoff) is the way in which the 
recombination remainder $\alpha_{\mathrm{rem}}$ is distributed among $\ell$ values at 
$\nmax$. In particular, there is more than one reasonable approach to topoff, and these
different approaches may lead to differences in the emission coefficients of certain lines.

These issues can be examined by comparing the results of the present calculation
with those of a second independent non-$J$-resolved calculation, Cloudy 
(see Appendix~\ref{Appdx:J} and Ferland et al. 1998). The approach to topoff used in Cloudy
differs somewhat from that described in Section~\ref{topoff}. 
The $J$-resolved calculation distributes $\alpha_{\mathrm{rem}}$ according to statistical weights,
while Cloudy assumes the levels are populated according to statistical weight.
These would be equivalent if the inverse lifetimes of these levels were proportional to statistical weight,
which they are not.

Both calculations are evaluated twice, with and without topoff. 
Figure~\ref{fig2} displays the emission coefficients of several 
strong optical and infrared lines, in each of the four cases, 
as a function of $\nmax$. Topoff is included in the top two panels but
not in the bottom two. The left two panels show the results of Cloudy,
and the right two show the results of the $J$-resolved calculations.
We normalize each emission coefficient to the average emission 
coefficient at $\nmax=100$. In each panel, the four lines bearing 
symbols designate cases that exhibit the greatest disagreement 
or slowest convergence with increasing $\nmax$.

With topoff included, Cloudy converges more rapidly
than the $J$-resolved code, a result of differing implementations of topoff. 
For most of the lines plotted, the difference 
between the two codes at $\nmax=100$ is less than $1\%$.
With topoff not included, most lines again agree to better than $1\%$,
although there are also significant outliers.  The lines bearing
symbols originate from yrast levels and their near neighbors.
These levels are most affected by the inclusion of topoff
and its method of implementation,  due to the restrictive selection rules that 
govern their decays. An yrast level (with $l = n-1$) can only decay to one
other yrast level (with $n^\prime = n - 1$ and $l^\prime = n^\prime - 1$) or to the
level $n^\prime = n$, $l^\prime = n - 2$ via a $\Delta n = 0$ transition. 
The yrast-to-yrast decay is far more likely than the $\Delta n = 0$ decay.
Thus, an yrast-to-yrast decay most likely will be followed by another yrast-to-yrast decay. 
It follows that any fraction of the recombination remainder,
$\alpha_{\mathrm{rem}}$, added to the yrast level at $\nmax$ increases the
effective recombination of all lower yrast levels by nearly the same amount.
Thus, the effects of including topoff are not yet negligible even at $\nmax=100$ for yrast levels. 
However, in a real atom  at finite densities, collisions will dominate (Porter {\it et~al.}~2005)
the very highest $n$-levels and force the populations into LTE.

\subsection{Effects of uncertainties in the atomic data}
\begin{figure*}[htbp]
\includegraphics*[angle=-90,width=6.5in,keepaspectratio=true]{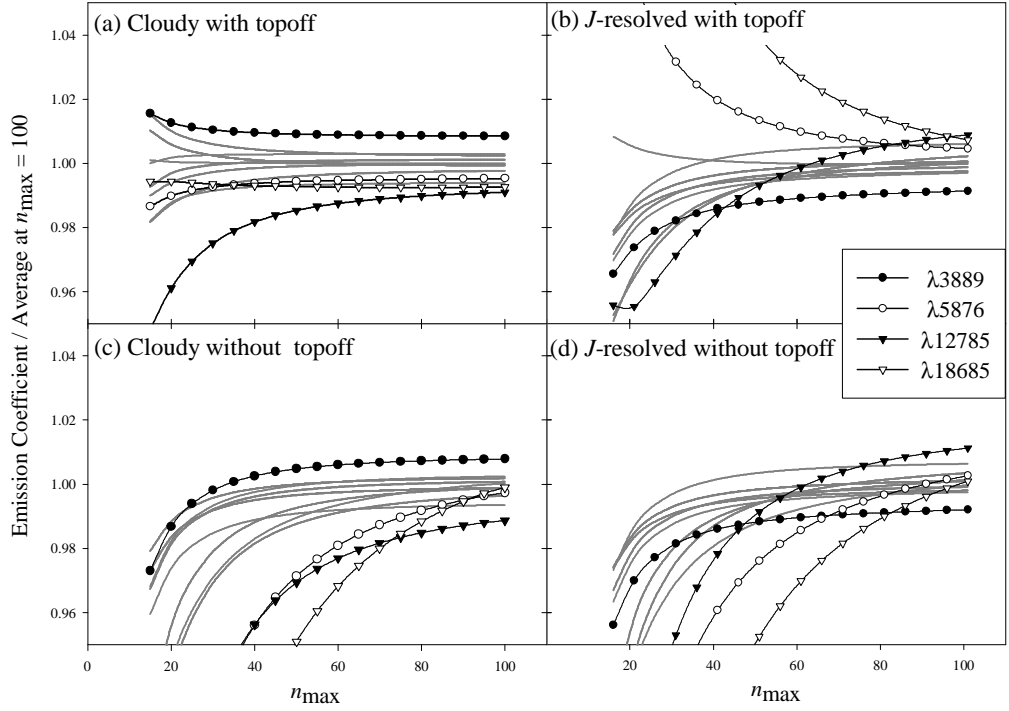}
\caption[]
{
The effects of increasing $\nmax$ on the convergence of emission coefficients is shown. 
The emission coefficients are results from the two different model calculations with and without topoff as follows:
a)~Cloudy (Ferland et al. 1998) with topoff; 
b)~$J$-resolved with topoff;
c)~Cloudy without topoff; and 
d)~$J$-resolved without topoff.
The average emission coefficient used to normalize the results is the average of the two model calculations at $\nmax=100$.
}
\label{fig2}
\end{figure*}

The lower two panels of Figure~\ref{fig2} show the two calculations without topoff out 
to $\nmax=100$. The majority of lines shown in the two lower panels of Figure~\ref{fig2}  
appear to have converged and show agreement to better than $1.0\%$.
However, lines from yrast-to-yrast 
transitions (indicated by symbols in the figure) appear not to have converged for $\nmax=100$.
For the lines which have converged, the differences are entirely due to 
the atomic data. There exist gaps in the atomic data that must be bridged, between the 
region where exact accurate variational results exist and the region where the hydrogenic approximation 
becomes applicable. The two codes use different reasonable approximations to bridge these 
gaps, and this introduces an uncertainty which we quantify here.

The Einstein $A$ coefficients introduce the lesser degree of uncertainty. Transitions between 
high-angular momentum levels are hydrogenic to a sufficient degree of accuracy. Transitions involving
$S$, $P$,~and $D$ levels involve different approximations, including semi-classical quantum defects and 
extrapolation from low-$n$ data.

Recombination coefficients, which are derived from photoionization cross sections, are the
greater source of uncertainty. Cross sections for 
$10 \le n \le 25$ and $L \le 2$  are the least accurate of these.

\subsection{Emission coefficients of representative He lines}

Table~\ref{table:Confidence} presents multiplet emission coefficients for lines satisfying the following criteria:
$n_{u} \le 15$, $\lambda  < 100\ \mu\mathrm{m}$, 
and 
$j_{\lambda }/j_{10830} \ge 10^{ - 3}$.
Each emission coefficient is the average, with $\nmax=100$, of the results from Cloudy and the $J$-resolved code,
with the individual fine-structure components in the $J$-resolved calculation summed.  Column~4 gives these average
emission coefficients without topoff, and Column~5 gives confidence estimates based on the differences between Cloudy
and the $J$-resolved code, again without topoff.  Columns~7 and 8 respectively present these values with topoff included.  
Confidence symbols correspond to percent difference between the results of Cloudy and the $J$-resolved code:
AA, A, B, and C signify that the results differ by less than $0.1\%$, less than $1.0\%$, less than $5.0\%$,
and more than $5.0\%$, respectively. Column~6 is the percent difference between columns~4 and 7.  

In Table~\ref{table:BSS-Comp} we present our final values along with the
lowest density $(n_{e} = 100 \mathrm{~cm}^{ - 3 })$ case of~\citetalias{BSS99}.  The small but
unknown collisional contributions to the results of~\citetalias{BSS99} prevent a rigorous comparison.  
Some transitions may also differ by a few percent because BSS99 did not scale the TOPbase photoionization
cross sections to agree with accurate ab initio cross sections at threshold.

\section{Conclusions}
\label{Conc}
We reach the following conclusions:

\begin{description}
\item[] A definitive test for the helium abundance produced in the Big Bang (Olive \& Skillman 2004)
requires that its abundance be measured to an accuracy of better than $1/2\%$. The requirement
for the He~I emission coefficients are similar. Several of the most important lines calculated here do not
meet that accuracy requirement. 

\item[] Improvements in the atomic data will be required to achieve that accuracy.
Our final accuracy is limited by gaps in the atomic data, mainly 
photoionization cross sections for intermediate-$n$, low-$L$ levels. 
An extension of the bound-bound oscillator strengths for low-$L$ transitions
will also improve further recombination-cascade calculations.

\item[] Singlet-triplet mixing does not affect intensities of multiplets, although 
intensities of lines within a multiplet can be strongly affected.  There may be an effect
at finite densities or with realistic radiative transfer.

\item[] Multiplets are not resolved in most astronomical sources since the intrinsic 
line widths are greater than the line splittings. It is not necessary to 
resolve fine structure in future calculations of the He~I emission spectrum.

\item[] In the low-density limit there is an additional uncertainty introduced by 
the need to ``top off'' a finite numerical representation of the 
infinite-level atom. This uncertainty can amount to $1\%$ for 
yrast-to-yrast lines but will not occur in actual nebulae. 
These have densities high enough for collisional 
processes to force populations of very highly excited levels into statistical equilibrium.

\item[] The predictions in Table~\ref{table:Confidence} (columns~7~\&~8) can be used to identify those lines 
that are least affected by gaps in the atomic data. These lines 
should be used when precise helium abundances are the desired end product.
\end{description}

Both of the codes discussed here are freely available and open source.  Cloudy can be downloaded from 
http://www.nublado.org, and the $J$-resolved code can be found at http://www.pa.uky.edu/$\sim$rporter.

We thank the NSF and NASA for support of this project through AST~03~07720 
and NAG5-12020 and UK's Center for Computational Sciences for a generous 
allocation of computer time.  We also thank referee Peter Storey for his helpful suggestions.

\appendix
\section{The non-$J$-resolved treatment in Cloudy}
\label{Appdx:J}
The recombination problem in the non-$J$-resolved code Cloudy (Ferland {\it et~al.}~1998) was solved as follows:

Energies for levels not included in the calculations of~\citetalias{Drake96} are calculated by assuming constant quantum defects
for $n \ge 10$. For levels with $L \ge 8$, the quantum defects are calculated
from a power law extrapolation of the lower $L$ defects at $ n = 10$. These differences are by far the most
accurately known and the most consistent between Cloudy and the $J$-resolved code. In both 
calculations there is essentially no uncertainty due to energies.

Emission oscillator strengths for $n_u \ge 11$, not included in the calculations of~\citetalias{Drake96},
are calculated by the extrapolation method outlined in Section~\ref{sec:extrapolation}
for transitions with $n_u \ge 11$, $n_l \le 5$, and both $L_u$ and $L_l \le 2$.
Emission oscillator strengths for hydrogenic transitions with $n_u \ge 11$, $n_l < n_u$,
and both $L_u$ and $L_l \ge 2$, are calculated by the method of HB90 discussed in 
Section~\ref{sec:Hoang-Bing}. All other oscillator strengths are calculated using
the semi-classical quantum defect method of Drake (1996, Chapter 7). The probability
for the forbidden transition $2 {}^1P-2 {}^3S$ is from Lach \& Pachucki~(2001). The most significant
discrepancies (and uncertainties) in oscillator strengths between Cloudy and the
$J$-resolved code are for levels with $n_u\ge11$, $n_l>5$, and both $L_l$ and $L_u<2$. 

We use fits to the TOPbase photoionization cross sections for the following levels:
$n {}^{1,3}S$ for $n\le 10$; $2 {}^3P$ and $3 {}^3P$; and $n^1$P for $n\le 7$.
P67 is used for the following levels: $n {}^3P$ for $4 \le n \le 10$; and $n {}^{1,3}D$ for $n \le 10$.
All other cross sections are calculated using a scaled hydrogenic method as
in Section~\ref{sec:hydro-cs}. Cross sections for levels with $n\le 4$ are
rescaled to agree at threshold with the \textit{ab initio} values calculated by \citetalias{HS98}.
For levels with $n = 5$ they are rescaled to values computed by the extrapolation method outlined by \citetalias{HS98}.
Differences in photoionization cross sections between our two codes are most significant for
levels with $L\le 2$, while cross sections for levels with $L > 2$ are essentially identical
and have negligible uncertainties. Photoionization cross sections, and by extension
recombination coefficients, are the greatest uncertainties in our calculations.

Cloudy treats topoff differently from the $J$-resolved code. Cloudy employs a ``collapsed''
level at $\nmax$ in which all of the individual $nLS$ terms are brought together
as one pseudo-level. The recombination coefficient into this pseudo-level is the sum
of recombination coefficients into the individual terms (calculated as in Section~\ref{sec:rrr},
with the changes in photoionization cross sections noted above) plus the recombination
remainder. Transition probabilities from this pseudo-level are calculated as follows
\begin{equation}
A(\nmax\rightarrow n_l,L_l,S) = \frac{ \sum\limits_{L_u=L_l\pm 1} {g_{L_u,S}A(\nmax,L_u,S\rightarrow n_l,L_l,S)} }
{ \sum\limits_{L_u=L_l\pm 1}{g_{L_u,S}} } 
\end{equation}
This causes the collapsed level to behave exactly as if it were a set of resolved terms populated according
to statistical weight.

\clearpage
\thispagestyle{empty}
\input{tab1.tex}
\clearpage
\input{tab2.tex} 
\clearpage
\input{tab3.tex}
\clearpage
\input{tab4.tex}

\clearpage

\clearpage

\end{document}

%% file: tab1.tex
\begin{table*}[tbhp]
\begin{center}
\caption{
The method used to calculated the oscillator strengths depends on the upper and lower levels. 
KEY: 
A)~\citetalias{Drake96} 
B)~extrapolation of~\citetalias{Drake96} 
C)~\citetalias{vanR}
D)~hydrogenic 
E)~various including non-dipole transitions 
X)~dipole transitions not included in Case~B.
The method used is independent of $S$ and $J$ except for transitions labeled with E as these will include non-dipole transitions.
\label{table:OscType}
}
\usefont{OT1}{cmtt}{m}{n}
\begin{tabular}{|rrr|r|r|r|r|r|r|r|r|r|r|r|r|}
\hline
n  &   & & 2&3  &4   &5    &6     &7      &8       &9        &10        &11         &12          &13           \\
\hline                                                                                                                          
   & L & &01&012&0123&01234&012345&0123456&01234567&012345678&0123456789&01234567891&012345678911&0123456789111\\
   &   & &  &   &    &     &      &       &        &         &          &          0&          01&          012\\
\hline                                                                                                  
\multicolumn{3}{|c|}{lower}&&&&&  &       &        &         &          &           &            &             \\
\hline                                                                                                                
 1 & 0 & &EE&.X.&.X..&.X...&.X....&.X.....&.X......&.X.......&.X........&.X.........&.X..........&.X...........\\
\hline                                                                                                                
 2 & 0 & &.E&.A.&.A..&.A...&.A....&.A.....&.A......&.A.......&.A........&.B.........&.B..........&.B...........\\
   & 1 & &..&A.A&A.A.&A.A..&A.A...&A.A....&A.A.....&A.A......&A.A.......&B.B........&B.B.........&B.B..........\\
\hline                                                                                                                          
 3 & 0 & &..&.A.&.A..&.A...&.A....&.A.....&.A......&.A.......&.A........&.B.........&.B..........&.B...........\\
   & 1 & &..&...&A.A.&A.A..&A.A...&A.A....&A.A.....&A.A......&A.A.......&B.B........&B.B.........&B.B..........\\
   & 2 & &..&.A.&.A.A&.A.A.&.A.A..&.A.A...&.A.A....&.A.A.....&.A.A......&.B.B.......&.B.B........&.B.B.........\\
\hline                                                                                                                          
 4 & 0 & &..&...&.A..&.A...&.A....&.A.....&.A......&.A.......&.A........&.B.........&.B..........&.B...........\\
   & 1 & &..&...&....&A.A..&A.A...&A.A....&A.A.....&A.A......&A.A.......&B.B........&B.B.........&B.B..........\\
   & 2 & &..&...&.A.A&.A.A.&.A.A..&.A.A...&.A.A....&.A.A.....&.A.A......&.B.B.......&.B.B........&.B.B.........\\
   & 3 & &..&...&....&..A.A&..A.A.&..A.A..&..A.A...&..A.A....&..A.A.....&..B.B......&..B.B.......&..B.B........\\
\hline                                                                                                                          
 5 & 0 & &..&...&....&.A...&.A....&.A.....&.A......&.A.......&.A........&.C.........&.C..........&.C...........\\
   & 1 & &..&...&....&.....&A.A...&A.A....&A.A.....&A.A......&A.A.......&C.C........&C.C.........&C.C..........\\
   & 2 & &..&...&....&.A.A.&.A.A..&.A.A...&.A.A....&.A.A.....&.A.A......&.C.C.......&.C.C........&.C.C.........\\
   & 3 & &..&...&....&....A&..A.A.&..A.A..&..A.A...&..A.A....&..A.A.....&..C.C......&..C.C.......&..C.C........\\
   & 4 & &..&...&....&.....&...A.A&...A.A.&...A.A..&...A.A...&...A.A....&...D.D.....&...D.D......&...D.D.......\\
\hline                                                                                                                          
 6 & 0 & &..&...&....&.....&.A....&.A.....&.A......&.A.......&.A........&.C.........&.C..........&.C...........\\
   & 1 & &..&...&....&.....&......&A.A....&A.A.....&A.A......&A.A.......&C.C........&C.C.........&C.C..........\\
   & 2 & &..&...&....&.....&.A.A..&.A.A...&.A.A....&.A.A.....&.A.A......&.C.C.......&.C.C........&.C.C.........\\
   & 3 & &..&...&....&.....&....A.&..A.A..&..A.A...&..A.A....&..A.A.....&..C.C......&..C.C.......&..C.C........\\
   & 4 & &..&...&....&.....&.....A&...A.A.&...A.A..&...A.A...&...A.A....&...C.C.....&...C.C......&...C.C.......\\
   & 5 & &..&...&....&.....&......&....A.A&....A.A.&....A.A..&....A.A...&....D.D....&....D.D.....&....D.D......\\
\hline                                                                                                                          
 7 & 0 & &..&...&....&.....&......&.A.....&.A......&.A.......&.A........&.C.........&.C..........&.C...........\\
   & 1 & &..&...&....&.....&......&.......&A.A.....&A.A......&A.A.......&C.C........&C.C.........&C.C..........\\
   & 2 & &..&...&....&.....&......&.A.A...&.A.A....&.A.A.....&.A.A......&.C.C.......&.C.C........&.C.C.........\\
   & 3 & &..&...&....&.....&......&....A..&..A.A...&..A.A....&..A.A.....&..C.C......&..C.C.......&..C.C........\\
   & 4 & &..&...&....&.....&......&.....A.&...A.A..&...A.A...&...A.A....&...C.C.....&...C.C......&...C.C.......\\
   & 5 & &..&...&....&.....&......&......A&....A.A.&....A.A..&....A.A...&....C.C....&....C.C.....&....C.C......\\
   & 6 & &..&...&....&.....&......&.......&.....A.D&.....A.D.&.....A.D..&.....D.D...&.....D.D....&.....D.D.....\\
\hline                                                                                                                          
 8 & 0 & &..&...&....&.....&......&.......&.A......&.A.......&.A........&.C.........&.C..........&.C...........\\
   & 1 & &..&...&....&.....&......&.......&........&A.A......&A.A.......&C.C........&C.C.........&C.C..........\\
   & 2 & &..&...&....&.....&......&.......&.A.A....&.A.A.....&.A.A......&.C.C.......&.C.C........&.C.C.........\\
   & 3 & &..&...&....&.....&......&.......&....A...&..A.A....&..A.A.....&..C.C......&..C.C.......&..C.C........\\
   & 4 & &..&...&....&.....&......&.......&.....A..&...A.A...&...A.A....&...C.C.....&...C.C......&...C.C.......\\
   & 5 & &..&...&....&.....&......&.......&......A.&....A.A..&....A.A...&....C.C....&....C.C.....&....C.C......\\
   & 6 & &..&...&....&.....&......&.......&.......C&.....A.D.&.....A.D..&.....C.D...&.....C.D....&.....C.D.....\\
   & 7 & &..&...&....&.....&......&.......&........&......A.D&......A.D.&......D.D..&......D.D...&......D.D....\\
\hline                                                                                                                          
 9 & 0 & &..&...&....&.....&......&.......&........&.A.......&.A........&.C.........&.C..........&.C...........\\
   & 1 & &..&...&....&.....&......&.......&........&.........&A.A.......&C.C........&C.C.........&C.C..........\\
   & 2 & &..&...&....&.....&......&.......&........&.A.A.....&.A.A......&.C.C.......&.C.C........&.C.C.........\\
   & 3 & &..&...&....&.....&......&.......&........&....A....&..A.A.....&..C.C......&..C.C.......&..C.C........\\
   & 4 & &..&...&....&.....&......&.......&........&.....A...&...A.A....&...C.C.....&...C.C......&...C.C.......\\
   & 5 & &..&...&....&.....&......&.......&........&......A..&....A.A...&....C.C....&....C.C.....&....C.C......\\
   & 6 & &..&...&....&.....&......&.......&........&.......C.&.....A.D..&.....C.D...&.....C.D....&.....C.D.....\\
   & 7 & &..&...&....&.....&......&.......&........&........C&......A.D.&......D.D..&......D.D...&......D.D....\\
 9 & 8 & &..&...&....&.....&......&.......&........&.........&.......D.D&.......D.D.&.......D.D..&.......D.D...\\
\hline                                                                                                                          
10 & 0 & &..&...&....&.....&......&.......&........&.........&.A........&.C.........&.C..........&.C...........\\
   & 1 & &..&...&....&.....&......&.......&........&.........&..........&C.C........&C.C.........&C.C..........\\
   & 2 & &..&...&....&.....&......&.......&........&.........&.A.A......&.C.C.......&.C.C........&.C.C.........\\
   & 3 & &..&...&....&.....&......&.......&........&.........&....A.....&..C.C......&..C.C.......&..C.C........\\
   & 4 & &..&...&....&.....&......&.......&........&.........&.....A....&...C.C.....&...C.C......&...C.C.......\\
   & 5 & &..&...&....&.....&......&.......&........&.........&......A...&....C.C....&....C.C.....&....C.C......\\
   & 6 & &..&...&....&.....&......&.......&........&.........&.......C..&.....C.D...&.....C.D....&.....C.D.....\\
   & 7 & &..&...&....&.....&......&.......&........&.........&........C.&......D.D..&......D.D...&......D.D....\\
   & 8 & &..&...&....&.....&......&.......&........&.........&.........C&.......D.D.&.......D.D..&.......D.D...\\
   & 9 & &..&...&....&.....&......&.......&........&.........&..........&........D.D&........D.D.&........D.D..\\
\hline                                                                                                                          
11 & 0 & &..&...&....&.....&......&.......&........&.........&..........&.C.........&.C..........&.C...........\\
   & 1 & &..&...&....&.....&......&.......&........&.........&..........&...........&C.C.........&C.C..........\\
   & 2 & &..&...&....&.....&......&.......&........&.........&..........&.C.C.......&.C.C........&.C.C.........\\
   & 3 & &..&...&....&.....&......&.......&........&.........&..........&....C......&..C.C.......&..C.C........\\
   & 4 & &..&...&....&.....&......&.......&........&.........&..........&.....C.....&...C.C......&...C.C.......\\
   & 5 & &..&...&....&.....&......&.......&........&.........&..........&......C....&....C.C.....&....C.C......\\
   & 6 & &..&...&....&.....&......&.......&........&.........&..........&.......C...&.....C.D....&.....C.D.....\\
   & 7 & &..&...&....&.....&......&.......&........&.........&..........&........C..&......D.D...&......D.D....\\
   & 8 & &..&...&....&.....&......&.......&........&.........&..........&.........C.&.......D.D..&.......D.D...\\
   & 9 & &..&...&....&.....&......&.......&........&.........&..........&..........C&........D.D.&........D.D..\\
   &10 & &..&...&....&.....&......&.......&........&.........&..........&...........&.........D.D&.........D.D.\\
\hline                                                                                                                          
12 & 0 & &..&...&....&.....&......&.......&........&.........&..........&...........&.C..........&.C...........\\
   & 1 & &..&...&....&.....&......&.......&........&.........&..........&...........&............&C.C..........\\
   & 2 & &..&...&....&.....&......&.......&........&.........&..........&...........&.C.C........&.C.C.........\\
   & 3 & &..&...&....&.....&......&.......&........&.........&..........&...........&....C.......&..C.C........\\
   & 4 & &..&...&....&.....&......&.......&........&.........&..........&...........&.....C......&...C.C.......\\
   & 5 & &..&...&....&.....&......&.......&........&.........&..........&...........&......C.....&....C.C......\\
   & 6 & &..&...&....&.....&......&.......&........&.........&..........&...........&.......C....&.....C.D.....\\
   & 7 & &..&...&....&.....&......&.......&........&.........&..........&...........&........C...&......D.D....\\
   & 8 & &..&...&....&.....&......&.......&........&.........&..........&...........&.........C..&.......D.D...\\
   & 9 & &..&...&....&.....&......&.......&........&.........&..........&...........&..........C.&........D.D..\\
   &10 & &..&...&....&.....&......&.......&........&.........&..........&...........&...........C&.........D.D.\\
   &11 & &..&...&....&.....&......&.......&........&.........&..........&...........&............&..........D.D\\
\hline                                                                                                                          
\end{tabular}
\end{center}
\end{table*}

%% file: tab2.tex
\begin{deluxetable}{clccc}
\tabletypesize{\scriptsize}
\tablecaption{
Comparison of emission coefficients, $4\pi j_{\lambda}/n_{e}n_{\sHe^+}$ 
for representative multiplets 
$3\,{}^{1,3}D$~--~$2\,{}^{1,3}P$ and 
$4\,{}^{1,3}F$~--~$3\,{}^{1,3}D$  
assuming pure $LS$-coupling and $ST$-mixing. The calculated emission coefficients 
of the component lines that comprise the above multiplets are given. 
The component line emission coefficients are summed to show the observable 
multiplet emission coefficients. Although small differences can be seen in 
the individual components, the multiplet sums are insensitive to $ST$-mixing.
These calculations where carried out with $\nmax = 100$.}
\tablewidth{0pt}
\tablehead
{
  \colhead{Wavelength (Air)}                      &
  \colhead{Transition}                            &
  \colhead{$LS$-coupling \par emiss. coeff.}      &
  \colhead{$ST$-mixing   \par emiss. coeff.}      &
  \colhead{Ratio of emiss. coeff. }              \\
  \colhead{$\mbox{\AA}$}                          &
  \colhead{}                                      &
  \colhead{$\mathrm{erg~cm}^{3}\mathrm{~s}^{-1}$} &
  \colhead{$\mathrm{erg~cm}^{3}\mathrm{~s}^{-1}$} &
  \colhead{$LS$-coupling/$ST$-mixing }           \\
}
\startdata
\textbf{5874.456} & $3 \,{}^{1}D_{2} $ -- $2 \,{}^{3}P_{2} $ &  ---             &  3.625369E-30  & ---    \\
\textbf{5874.483} & $3 \,{}^{1}D_{2} $ -- $2 \,{}^{3}P_{1} $ &  ---             &  1.036918E-29  & ---    \\
\textbf{5875.621} & $3 \,{}^{3}D_{1} $ -- $2 \,{}^{3}P_{2} $ &  9.215998E-28    &  9.216010E-28  & 1.0000 \\
\textbf{5875.636} & $3 \,{}^{3}D_{2} $ -- $2 \,{}^{3}P_{2} $ &  1.382263E-26    &  1.382984E-26  & 0.9995 \\
\textbf{5875.637} & $3 \,{}^{3}D_{3} $ -- $2 \,{}^{3}P_{2} $ &  7.740541E-26    &  7.741031E-26  & 0.9999 \\
\textbf{5875.648} & $3 \,{}^{3}D_{1} $ -- $2 \,{}^{3}P_{1} $ &  1.382384E-26    &  1.382385E-26  & 1.0000 \\
\textbf{5875.663} & $3 \,{}^{3}D_{2} $ -- $2 \,{}^{3}P_{1} $ &  4.146730E-26    &  4.148941E-26  & 0.9995 \\
\textbf{5875.989} & $3 \,{}^{3}D_{1} $ -- $2 \,{}^{3}P_{0} $ &  1.842857E-26    &  1.842860E-26  & 1.0000 \\
\textbf{Sum}      & $3 \,{}^{3}D_{ } $ -- $2 \,{}^{3}P_{ } $ &  1.658693E-25    &  1.659176E-25  & 0.9997 \\
                  &                                          &                  &                &        \\
\textbf{6678.180} & $3 \,{}^{1}D_{2} $ -- $2 \,{}^{1}P_{1} $ &  4.713582E-26    &  4.713351E-26  & 1.0000 \\
\textbf{6679.686} & $3 \,{}^{3}D_{1} $ -- $2 \,{}^{1}P_{1} $ &  ---             &  7.287124E-34  & ---    \\
\textbf{6679.705} & $3 \,{}^{3}D_{2} $ -- $2 \,{}^{1}P_{1} $ &  ---             &  1.039168E-29  & ---    \\
\textbf{Sum}      & $3 \,{}^{1}D_{ } $ -- $2 \,{}^{1}P_{ } $ &  4.713582E-26    &  4.714390E-26  & 0.9998 \\
                  &                                          &                  &                &        \\
\textbf{18685.14} & $4 \,{}^{1}F_{3} $ -- $3 \,{}^{3}D_{3} $ &  ---             &  3.080526E-28  & ---    \\
\textbf{18685.15} & $4 \,{}^{1}F_{3} $ -- $3 \,{}^{3}D_{2} $ &  ---             &  2.357295E-27  & ---    \\
\textbf{18685.17} & $4 \,{}^{3}F_{2} $ -- $3 \,{}^{3}D_{3} $ &  2.410942E-29    &  2.410945E-29  & 1.0000 \\
\textbf{18685.18} & $4 \,{}^{3}F_{2} $ -- $3 \,{}^{3}D_{2} $ &  8.438345E-28    &  8.436298E-28  & 1.0002 \\
\textbf{18685.20} & $4 \,{}^{3}F_{4} $ -- $3 \,{}^{3}D_{3} $ &  9.762777E-27    &  9.762664E-27  & 1.0000 \\
\textbf{18685.23} & $4 \,{}^{3}F_{3} $ -- $3 \,{}^{3}D_{3} $ &  8.437227E-28    &  5.359178E-28  & 1.5744 \\
\textbf{18685.23} & $4 \,{}^{3}F_{3} $ -- $3 \,{}^{3}D_{2} $ &  6.749771E-27    &  4.394646E-27  & 1.5359 \\
\textbf{18685.33} & $4 \,{}^{3}F_{2} $ -- $3 \,{}^{3}D_{1} $ &  4.556594E-27    &  4.556599E-27  & 1.0000 \\
\textbf{Sum}      & $4 \,{}^{3}F_{ } $ -- $3 \,{}^{3}D_{ } $ &  2.278081E-26    &  2.278291E-26  & 0.9999 \\
                  &                                          &                  &                &        \\
\textbf{18697.10} & $4 \,{}^{1}F_{3} $ -- $3 \,{}^{1}D_{2} $ &  7.589393E-27    &  4.925742E-27  & 1.5408 \\
\textbf{18697.12} & $4 \,{}^{3}F_{2} $ -- $3 \,{}^{1}D_{2} $ &  ---             &  2.052122E-31  & ---    \\
\textbf{18697.18} & $4 \,{}^{3}F_{3} $ -- $3 \,{}^{1}D_{2} $ &  ---             &  2.661372E-27  & ---    \\
\textbf{Sum}      & $4 \,{}^{1}F_{ } $ -- $3 \,{}^{1}D_{ } $ &  7.589393E-27    &  7.587319E-27  & 1.0003 \\
\enddata
\label{table:J-Results}
\end{deluxetable}

%% file: tab3.tex
\begin{deluxetable}{rrrccccc}
\tabletypesize{\scriptsize}
\tablecaption{
Average emission coefficients, $4\pi j_{\lambda}/n_{e}n_{\sHe^+}$ in lines meeting the simultaneous criteria: 
$n_{u} \le  15$, 
$\lambda  < 100\ \mu$m, 
and 
$j_{\lambda }/j_{10830} \ge  10^{ - 3}$. 
The confidence codes indicate the percent difference between the results of the two models: 
AA, A, B, and C correspond to a difference of 
less than 0.1{\%}, 
less than 1.0{\%}, 
less than 5.0{\%}, and 
more than 5.0{\%}, respectively. 
}
\tablewidth{0pt}
\tablehead{
        \colhead{}                                      & 
        \colhead{}                                      & 
        \colhead{}                                      & 
        \colhead{Emiss. coeff.}                         & 
        \colhead{}                                      & 
        \colhead{}                                      & 
        \colhead{Emiss. coeff.}                         & 
        \colhead{}                                      \\
        \colhead{Wavelength (Air)}                      & 
        \colhead{$n\,{}^{2S+1}\! L $}                   & 
        \colhead{$n\,{}^{2S+1}\! L $}                   & 
        \colhead{``no~topoff"}                          & 
        \colhead{Confidence}                            & 
        \colhead{\% diff.}                              & 
        \colhead{``topoff"}                             & 
        \colhead{Confidence}                            \\
        \colhead{$\mbox{\AA}$}                          & 
        \colhead{upper}                                 & 
        \colhead{lower}                                 & 
        \colhead{$\mathrm{erg~cm}^{3}\mathrm{~s}^{-1}$} & 
        \colhead{}                                      & 
        \colhead{}                                      & 
        \colhead{$\mathrm{erg~cm}^{3}\mathrm{~s}^{-1}$} & 
        \colhead{}                                      \\
}
\startdata
2633    &       $15\;{}^{3}P$   &       $2\;{}^{3}S$    &       8.907E-28       &       B       &       0.05{\%}        &       8.911E-28       &       B       \\
2638    &       $14\;{}^{3}P$   &       $2\;{}^{3}S$    &       1.095E-27       &       B       &       0.05{\%}        &       1.096E-27       &       B       \\
2645    &       $13\;{}^{3}P$   &       $2\;{}^{3}S$    &       1.368E-27       &       B       &       0.05{\%}        &       1.369E-27       &       B       \\
2653    &       $12\;{}^{3}P$   &       $2\;{}^{3}S$    &       1.740E-27       &       B       &       0.05{\%}        &       1.741E-27       &       B       \\
2663    &       $11\;{}^{3}P$   &       $2\;{}^{3}S$    &       2.263E-27       &       B       &       0.05{\%}        &       2.264E-27       &       B       \\
2677    &       $10\;{}^{3}P$   &       $2\;{}^{3}S$    &       3.027E-27       &       B       &       0.04{\%}        &       3.029E-27       &       B       \\
2696    &       $ 9\;{}^{3}P$   &       $2\;{}^{3}S$    &       4.168E-27       &       A       &       0.04{\%}        &       4.170E-27       &       AA      \\
2723    &       $ 8\;{}^{3}P$   &       $2\;{}^{3}S$    &       6.036E-27       &       A       &       0.04{\%}        &       6.039E-27       &       AA      \\
2764    &       $ 7\;{}^{3}P$   &       $2\;{}^{3}S$    &       9.127E-27       &       A       &       0.04{\%}        &       9.131E-27       &       AA      \\
2829    &       $ 6\;{}^{3}P$   &       $2\;{}^{3}S$    &       1.487E-26       &       A       &       0.04{\%}        &       1.488E-26       &       A       \\
2945    &       $ 5\;{}^{3}P$   &       $2\;{}^{3}S$    &       2.655E-26       &       A       &       0.05{\%}        &       2.657E-26       &       A       \\
3176    &       $14\;{}^{1}P$   &       $2\;{}^{1}S$    &       2.944E-28       &       C       &       0.06{\%}        &       2.946E-28       &       B       \\
3185    &       $13\;{}^{1}P$   &       $2\;{}^{1}S$    &       3.689E-28       &       B       &       0.06{\%}        &       3.691E-28       &       B       \\
3188    &       $ 4\;{}^{3}P$   &       $2\;{}^{3}S$    &       5.561E-26       &       A       &       0.06{\%}        &       5.564E-26       &       A       \\
3197    &       $12\;{}^{1}P$   &       $2\;{}^{1}S$    &       4.703E-28       &       B       &       0.06{\%}        &       4.706E-28       &       B       \\
3212    &       $11\;{}^{1}P$   &       $2\;{}^{1}S$    &       6.121E-28       &       B       &       0.05{\%}        &       6.125E-28       &       B       \\
3231    &       $10\;{}^{1}P$   &       $2\;{}^{1}S$    &       8.170E-28       &       B       &       0.05{\%}        &       8.174E-28       &       B       \\
3258    &       $ 9\;{}^{1}P$   &       $2\;{}^{1}S$    &       1.119E-27       &       A       &       0.05{\%}        &       1.119E-27       &       A       \\
3297    &       $ 8\;{}^{1}P$   &       $2\;{}^{1}S$    &       1.607E-27       &       A       &       0.05{\%}        &       1.608E-27       &       A       \\
3355    &       $ 7\;{}^{1}P$   &       $2\;{}^{1}S$    &       2.409E-27       &       A       &       0.05{\%}        &       2.410E-27       &       A       \\
3448    &       $ 6\;{}^{1}P$   &       $2\;{}^{1}S$    &       3.869E-27       &       B       &       0.04{\%}        &       3.870E-27       &       A       \\
3479    &       $15\;{}^{3}D$   &       $2\;{}^{3}P$    &       9.696E-28       &       B       &       0.09{\%}        &       9.704E-28       &       B       \\
3488    &       $14\;{}^{3}D$   &       $2\;{}^{3}P$    &       1.193E-27       &       B       &       0.09{\%}        &       1.194E-27       &       B       \\
3499    &       $13\;{}^{3}D$   &       $2\;{}^{3}P$    &       1.490E-27       &       B       &       0.09{\%}        &       1.491E-27       &       B       \\
3513    &       $12\;{}^{3}D$   &       $2\;{}^{3}P$    &       1.896E-27       &       B       &       0.08{\%}        &       1.897E-27       &       B       \\
3531    &       $11\;{}^{3}D$   &       $2\;{}^{3}P$    &       2.465E-27       &       B       &       0.08{\%}        &       2.467E-27       &       B       \\
3554    &       $10\;{}^{3}D$   &       $2\;{}^{3}P$    &       3.324E-27       &       B       &       0.08{\%}        &       3.327E-27       &       B       \\
3587    &       $ 9\;{}^{3}D$   &       $2\;{}^{3}P$    &       4.575E-27       &       A       &       0.08{\%}        &       4.579E-27       &       A       \\
3599    &       $ 9\;{}^{3}S$   &       $2\;{}^{3}P$    &       3.057E-28       &       C       &       0.02{\%}        &       3.058E-28       &       C       \\
3614    &       $ 5\;{}^{1}P$   &       $2\;{}^{1}S$    &       6.859E-27       &       A       &       0.06{\%}        &       6.864E-27       &       A       \\
3634    &       $ 8\;{}^{3}D$   &       $2\;{}^{3}P$    &       6.574E-27       &       A       &       0.08{\%}        &       6.579E-27       &       A       \\
3652    &       $ 8\;{}^{3}S$   &       $2\;{}^{3}P$    &       4.538E-28       &       C       &       0.02{\%}        &       4.539E-28       &       C       \\
3705    &       $ 7\;{}^{3}D$   &       $2\;{}^{3}P$    &       9.934E-27       &       A       &       0.08{\%}        &       9.942E-27       &       A       \\
3733    &       $ 7\;{}^{3}S$   &       $2\;{}^{3}P$    &       7.290E-28       &       B       &       0.03{\%}        &       7.292E-28       &       B       \\
3756    &       $14\;{}^{1}D$   &       $2\;{}^{1}P$    &       3.141E-28       &       A       &       0.09{\%}        &       3.144E-28       &       A       \\
3769    &       $13\;{}^{1}D$   &       $2\;{}^{1}P$    &       3.924E-28       &       A       &       0.09{\%}        &       3.928E-28       &       A       \\
3785    &       $12\;{}^{1}D$   &       $2\;{}^{1}P$    &       4.993E-28       &       A       &       0.09{\%}        &       4.997E-28       &       A       \\
3806    &       $11\;{}^{1}D$   &       $2\;{}^{1}P$    &       6.491E-28       &       A       &       0.09{\%}        &       6.497E-28       &       A       \\
3820    &       $ 6\;{}^{3}D$   &       $2\;{}^{3}P$    &       1.613E-26       &       A       &       0.09{\%}        &       1.614E-26       &       A       \\
3834    &       $10\;{}^{1}D$   &       $2\;{}^{1}P$    &       8.577E-28       &       B       &       0.09{\%}        &       8.584E-28       &       B       \\
3868    &       $ 6\;{}^{3}S$   &       $2\;{}^{3}P$    &       1.263E-27       &       B       &       0.03{\%}        &       1.263E-27       &       B       \\
3872    &       $ 9\;{}^{1}D$   &       $2\;{}^{1}P$    &       1.182E-27       &       A       &       0.09{\%}        &       1.183E-27       &       A       \\
3889    &       $ 3\;{}^{3}P$   &       $2\;{}^{3}S$    &       1.380E-25       &       B       &       0.07{\%}        &       1.381E-25       &       B       \\
3927    &       $ 8\;{}^{1}D$   &       $2\;{}^{1}P$    &       1.704E-27       &       A       &       0.09{\%}        &       1.705E-27       &       A       \\
3965    &       $ 4\;{}^{1}P$   &       $2\;{}^{1}S$    &       1.397E-26       &       A       &       0.06{\%}        &       1.398E-26       &       A       \\
4009    &       $ 7\;{}^{1}D$   &       $2\;{}^{1}P$    &       2.585E-27       &       A       &       0.09{\%}        &       2.587E-27       &       A       \\
4024    &       $ 7\;{}^{1}S$   &       $2\;{}^{1}P$    &       3.065E-28       &       A       &       0.05{\%}        &       3.067E-28       &       A       \\
4026    &       $ 5\;{}^{3}D$   &       $2\;{}^{3}P$    &       2.898E-26       &       A       &       0.10{\%}        &       2.901E-26       &       AA      \\
4121    &       $ 5\;{}^{3}S$   &       $2\;{}^{3}P$    &       2.490E-27       &       B       &       0.04{\%}        &       2.491E-27       &       B       \\
4144    &       $ 6\;{}^{1}D$   &       $2\;{}^{1}P$    &       4.225E-27       &       A       &       0.09{\%}        &       4.229E-27       &       A       \\
4169    &       $ 6\;{}^{1}S$   &       $2\;{}^{1}P$    &       5.212E-28       &       A       &       0.05{\%}        &       5.215E-28       &       A       \\
4388    &       $ 5\;{}^{1}D$   &       $2\;{}^{1}P$    &       7.669E-27       &       A       &       0.10{\%}        &       7.677E-27       &       A       \\
4438    &       $ 5\;{}^{1}S$   &       $2\;{}^{1}P$    &       1.003E-27       &       A       &       0.04{\%}        &       1.004E-27       &       A       \\
4472    &       $ 4\;{}^{3}D$   &       $2\;{}^{3}P$    &       6.102E-26       &       A       &       0.13{\%}        &       6.110E-26       &       A       \\
4713    &       $ 4\;{}^{3}S$   &       $2\;{}^{3}P$    &       6.426E-27       &       A       &       0.04{\%}        &       6.429E-27       &       A       \\
4922    &       $ 4\;{}^{1}D$   &       $2\;{}^{1}P$    &       1.649E-26       &       A       &       0.14{\%}        &       1.651E-26       &       A       \\
5016    &       $ 3\;{}^{1}P$   &       $2\;{}^{1}S$    &       3.506E-26       &       A       &       0.08{\%}        &       3.508E-26       &       A       \\
5048    &       $ 4\;{}^{1}S$   &       $2\;{}^{1}P$    &       2.416E-27       &       A       &       0.04{\%}        &       2.417E-27       &       A       \\
5876    &       $ 3\;{}^{3}D$   &       $2\;{}^{3}P$    &       1.627E-25       &       A       &       1.47{\%}        &       1.651E-25       &       A       \\
6678    &       $ 3\;{}^{1}D$   &       $2\;{}^{1}P$    &       4.620E-26       &       A       &       1.51{\%}        &       4.691E-26       &       A       \\
7065    &       $ 3\;{}^{3}S$   &       $2\;{}^{3}P$    &       2.866E-26       &       A       &       0.05{\%}        &       2.867E-26       &       A       \\
7281    &       $ 3\;{}^{1}S$   &       $2\;{}^{1}P$    &       8.712E-27       &       A       &       0.05{\%}        &       8.716E-27       &       A       \\
7298    &       $ 9\;{}^{3}P$   &       $3\;{}^{3}S$    &       3.301E-28       &       A       &       0.04{\%}        &       3.303E-28       &       AA      \\
7500    &       $ 8\;{}^{3}P$   &       $3\;{}^{3}S$    &       4.627E-28       &       A       &       0.04{\%}        &       4.629E-28       &       AA      \\
7816    &       $ 7\;{}^{3}P$   &       $3\;{}^{3}S$    &       6.644E-28       &       A       &       0.04{\%}        &       6.647E-28       &       AA      \\
8362    &       $ 6\;{}^{3}P$   &       $3\;{}^{3}S$    &       9.894E-28       &       A       &       0.04{\%}        &       9.898E-28       &       A       \\
8444    &       $11\;{}^{3}D$   &       $3\;{}^{3}P$    &       3.307E-28       &       B       &       0.08{\%}        &       3.309E-28       &       B       \\
8582    &       $14\;{}^{3}F$   &       $3\;{}^{3}D$    &       3.169E-28       &       A       &       0.14{\%}        &       3.174E-28       &       A       \\
8583    &       $10\;{}^{3}D$   &       $3\;{}^{3}P$    &       4.407E-28       &       B       &       0.08{\%}        &       4.410E-28       &       B       \\
8648    &       $13\;{}^{3}F$   &       $3\;{}^{3}D$    &       3.977E-28       &       A       &       0.13{\%}        &       3.982E-28       &       A       \\
8733    &       $12\;{}^{3}F$   &       $3\;{}^{3}D$    &       5.089E-28       &       A       &       0.13{\%}        &       5.096E-28       &       A       \\
8777    &       $ 9\;{}^{3}D$   &       $3\;{}^{3}P$    &       5.965E-28       &       A       &       0.08{\%}        &       5.970E-28       &       A       \\
8845    &       $11\;{}^{3}F$   &       $3\;{}^{3}D$    &       6.665E-28       &       A       &       0.14{\%}        &       6.674E-28       &       A       \\
8997    &       $10\;{}^{3}F$   &       $3\;{}^{3}D$    &       8.999E-28       &       AA      &       0.14{\%}        &       9.012E-28       &       A       \\
9000    &       $10\;{}^{1}F$   &       $3\;{}^{1}D$    &       3.000E-28       &       AA      &       0.14{\%}        &       3.004E-28       &       A       \\
9063    &       $ 8\;{}^{3}D$   &       $3\;{}^{3}P$    &       8.365E-28       &       A       &       0.08{\%}        &       8.372E-28       &       A       \\
9210    &       $ 9\;{}^{3}F$   &       $3\;{}^{3}D$    &       1.260E-27       &       A       &       0.14{\%}        &       1.262E-27       &       A       \\
9213    &       $ 9\;{}^{1}F$   &       $3\;{}^{1}D$    &       4.200E-28       &       A       &       0.14{\%}        &       4.206E-28       &       A       \\
9464    &       $ 5\;{}^{3}P$   &       $3\;{}^{3}S$    &       1.468E-27       &       A       &       0.05{\%}        &       1.469E-27       &       A       \\
9517    &       $ 7\;{}^{3}D$   &       $3\;{}^{3}P$    &       1.217E-27       &       A       &       0.08{\%}        &       1.218E-27       &       A       \\
9526    &       $ 8\;{}^{3}F$   &       $3\;{}^{3}D$    &       1.843E-27       &       A       &       0.15{\%}        &       1.846E-27       &       A       \\
9529    &       $ 8\;{}^{1}F$   &       $3\;{}^{1}D$    &       6.142E-28       &       A       &       0.15{\%}        &       6.151E-28       &       A       \\
9603    &       $ 6\;{}^{1}P$   &       $3\;{}^{1}S$    &       3.567E-28       &       B       &       0.04{\%}        &       3.569E-28       &       A       \\
10028   &       $ 7\;{}^{3}F$   &       $3\;{}^{3}D$    &       2.864E-27       &       A       &       0.16{\%}        &       2.869E-27       &       A       \\
10031   &       $ 7\;{}^{1}F$   &       $3\;{}^{1}D$    &       9.546E-28       &       A       &       0.16{\%}        &       9.561E-28       &       A       \\
10138   &       $ 7\;{}^{1}D$   &       $3\;{}^{1}P$    &       3.883E-28       &       A       &       0.09{\%}        &       3.887E-28       &       A       \\
10311   &       $ 6\;{}^{3}D$   &       $3\;{}^{3}P$    &       1.852E-27       &       A       &       0.09{\%}        &       1.854E-27       &       A       \\
10830   &       $ 2\;{}^{3}P$   &       $2\;{}^{3}S$    &       2.705E-25       &       AA      &       0.53{\%}        &       2.720E-25       &       AA      \\
10913   &       $ 6\;{}^{3}F$   &       $3\;{}^{3}D$    &       4.853E-27       &       A       &       0.18{\%}        &       4.862E-27       &       AA      \\
10917   &       $ 6\;{}^{1}F$   &       $3\;{}^{1}D$    &       1.617E-27       &       A       &       0.18{\%}        &       1.620E-27       &       AA      \\
10997   &       $ 6\;{}^{3}P$   &       $3\;{}^{3}D$    &       2.812E-28       &       A       &       0.04{\%}        &       2.813E-28       &       A       \\
11013   &       $ 5\;{}^{1}P$   &       $3\;{}^{1}S$    &       5.475E-28       &       A       &       0.06{\%}        &       5.479E-28       &       A       \\
11045   &       $ 6\;{}^{1}D$   &       $3\;{}^{1}P$    &       5.993E-28       &       A       &       0.09{\%}        &       5.999E-28       &       A       \\
11969   &       $ 5\;{}^{3}D$   &       $3\;{}^{3}P$    &       2.923E-27       &       A       &       0.10{\%}        &       2.926E-27       &       AA      \\
12527   &       $ 4\;{}^{3}P$   &       $3\;{}^{3}S$    &       1.781E-27       &       A       &       0.06{\%}        &       1.782E-27       &       A       \\
12785   &       $ 5\;{}^{3}F$   &       $3\;{}^{3}D$    &       9.454E-27       &       B       &       0.27{\%}        &       9.480E-27       &       B       \\
12790   &       $ 5\;{}^{1}F$   &       $3\;{}^{1}D$    &       3.150E-27       &       B       &       0.27{\%}        &       3.158E-27       &       B       \\
12846   &       $ 5\;{}^{3}S$   &       $3\;{}^{3}P$    &       4.900E-28       &       B       &       0.04{\%}        &       4.902E-28       &       B       \\
12968   &       $ 5\;{}^{1}D$   &       $3\;{}^{1}P$    &       9.704E-28       &       A       &       0.10{\%}        &       9.714E-28       &       A       \\
12985   &       $ 5\;{}^{3}P$   &       $3\;{}^{3}D$    &       5.135E-28       &       A       &       0.05{\%}        &       5.138E-28       &       A       \\
15084   &       $ 4\;{}^{1}P$   &       $3\;{}^{1}S$    &       7.424E-28       &       A       &       0.06{\%}        &       7.429E-28       &       A       \\
17002   &       $ 4\;{}^{3}D$   &       $3\;{}^{3}P$    &       4.315E-27       &       A       &       0.13{\%}        &       4.321E-27       &       A       \\
17330   &       $10\;{}^{3}F$   &       $4\;{}^{3}D$    &       2.863E-28       &       AA      &       0.14{\%}        &       2.867E-28       &       A       \\
17352   &       $10\;{}^{3}G$   &       $4\;{}^{3}F$    &       3.424E-28       &       AA      &       0.23{\%}        &       3.432E-28       &       A       \\
18139   &       $ 9\;{}^{3}F$   &       $4\;{}^{3}D$    &       3.914E-28       &       A       &       0.14{\%}        &       3.919E-28       &       A       \\
18163   &       $ 9\;{}^{3}G$   &       $4\;{}^{3}F$    &       4.922E-28       &       A       &       0.24{\%}        &       4.934E-28       &       A       \\
18685   &       $ 4\;{}^{3}F$   &       $3\;{}^{3}D$    &       2.190E-26       &       A       &       3.18{\%}        &       2.261E-26       &       B       \\
18697   &       $ 4\;{}^{1}F$   &       $3\;{}^{1}D$    &       7.296E-27       &       A       &       3.18{\%}        &       7.535E-27       &       B       \\
19089   &       $ 4\;{}^{1}D$   &       $3\;{}^{1}P$    &       1.523E-27       &       A       &       0.14{\%}        &       1.525E-27       &       A       \\
19406   &       $ 8\;{}^{3}F$   &       $4\;{}^{3}D$    &       5.515E-28       &       A       &       0.15{\%}        &       5.523E-28       &       A       \\
19434   &       $ 8\;{}^{3}G$   &       $4\;{}^{3}F$    &       7.446E-28       &       A       &       0.27{\%}        &       7.466E-28       &       AA      \\
19543   &       $ 4\;{}^{3}P$   &       $3\;{}^{3}D$    &       1.038E-27       &       A       &       0.06{\%}        &       1.039E-27       &       A       \\
21118   &       $ 4\;{}^{3}S$   &       $3\;{}^{3}P$    &       9.811E-28       &       A       &       0.04{\%}        &       9.815E-28       &       A       \\
21130   &       $ 4\;{}^{1}S$   &       $3\;{}^{1}P$    &       3.915E-28       &       A       &       0.04{\%}        &       3.917E-28       &       A       \\
21608   &       $ 7\;{}^{3}F$   &       $4\;{}^{3}D$    &       8.054E-28       &       A       &       0.16{\%}        &       8.067E-28       &       A       \\
21641   &       $ 7\;{}^{3}G$   &       $4\;{}^{3}F$    &       1.216E-27       &       A       &       0.33{\%}        &       1.220E-27       &       A       \\
21642   &       $ 7\;{}^{1}G$   &       $4\;{}^{1}F$    &       4.053E-28       &       A       &       0.33{\%}        &       4.066E-28       &       A       \\
24727   &       $ 6\;{}^{3}D$   &       $4\;{}^{3}P$    &       3.140E-28       &       A       &       0.09{\%}        &       3.143E-28       &       A       \\
26185   &       $ 6\;{}^{3}F$   &       $4\;{}^{3}D$    &       1.207E-27       &       A       &       0.18{\%}        &       1.210E-27       &       AA      \\
26198   &       $ 6\;{}^{1}F$   &       $4\;{}^{1}D$    &       4.028E-28       &       A       &       0.18{\%}        &       4.035E-28       &       AA      \\
26234   &       $ 6\;{}^{3}G$   &       $4\;{}^{3}F$    &       2.253E-27       &       B       &       0.54{\%}        &       2.265E-27       &       B       \\
26234   &       $ 6\;{}^{1}G$   &       $4\;{}^{1}F$    &       7.508E-28       &       B       &       0.54{\%}        &       7.549E-28       &       B       \\
37026   &       $ 5\;{}^{3}D$   &       $4\;{}^{3}P$    &       3.475E-28       &       A       &       0.10{\%}        &       3.479E-28       &       AA      \\
37372   &       $ 8\;{}^{3}G$   &       $5\;{}^{3}F$    &       3.330E-28       &       A       &       0.27{\%}        &       3.339E-28       &       AA      \\
37378   &       $ 8\;{}^{3}H$   &       $5\;{}^{3}G$    &       3.527E-28       &       A       &       0.59{\%}        &       3.548E-28       &       A       \\
40366   &       $ 5\;{}^{3}F$   &       $4\;{}^{3}D$    &       1.693E-27       &       B       &       0.27{\%}        &       1.697E-27       &       B       \\
40398   &       $ 5\;{}^{1}F$   &       $4\;{}^{1}D$    &       5.646E-28       &       B       &       0.27{\%}        &       5.662E-28       &       B       \\
40479   &       $ 5\;{}^{3}G$   &       $4\;{}^{3}F$    &       4.660E-27       &       A       &       6.36{\%}        &       4.976E-27       &       B       \\
40479   &       $ 5\;{}^{1}G$   &       $4\;{}^{1}F$    &       1.553E-27       &       A       &       6.36{\%}        &       1.658E-27       &       B       \\
42946   &       $ 3\;{}^{3}P$   &       $3\;{}^{3}S$    &       1.416E-27       &       B       &       0.07{\%}        &       1.417E-27       &       B       \\
46493   &       $ 7\;{}^{3}G$   &       $5\;{}^{3}F$    &       4.801E-28       &       A       &       0.33{\%}        &       4.817E-28       &       A       \\
46503   &       $ 7\;{}^{3}H$   &       $5\;{}^{3}G$    &       6.495E-28       &       C       &       1.03{\%}        &       6.562E-28       &       B       \\
74517   &       $ 6\;{}^{3}G$   &       $5\;{}^{3}F$    &       6.383E-28       &       B       &       0.54{\%}        &       6.418E-28       &       B       \\
74541   &       $ 6\;{}^{3}H$   &       $5\;{}^{3}G$    &       1.238E-27       &       B       &      11.73{\%}        &       1.403E-27       &       B       \\
74541   &       $ 6\;{}^{1}H$   &       $5\;{}^{1}G$    &       4.126E-28       &       B       &      11.74{\%}        &       4.675E-28       &       B       \\
123631  &       $ 7\;{}^{3}I$   &       $6\;{}^{3}H$    &       3.763E-28       &       C       &      20.21{\%}        &       4.716E-28       &       B       \\
\enddata
\label{table:Confidence}
\end{deluxetable}

%% file: tab4.tex
\begin{deluxetable}{rrrrrr}
\tabletypesize{\scriptsize}
\tablecaption{
Comparison of the present results with those of the lowest density ($100$ cm$^{-3}$)
case of~\citetalias{BSS99}. The~\citetalias{BSS99} results include collision contributions not considered in
this work (see text).
}
\tablewidth{0pt}
\tablehead
{
  \colhead{Wavelength (Air)}                               & 
  \colhead{$n \,{}^{2S+1}\! L$}                            & 
  \colhead{$n \,{}^{2S+1}\! L$}                            &
  \colhead{Present}                                        &
  \colhead{BSS99}                                          &
  \colhead{\% diff.}                                       \\
  \colhead{$\mbox{\AA}$}                                   &
  \colhead{upper}                                          &
  \colhead{lower}                                          &
  \colhead{$\mathrm{erg~cm}^{3}\mathrm{~s}^{-1}$}          &
  \colhead{$\mathrm{erg~cm}^{3}\mathrm{~s}^{-1}$}          &
  \colhead{}                                               \\
  \colhead{$\lambda$}                                      &
  \colhead{}                                               &
  \colhead{}                                               &
  \colhead{$4\pi j_{\lambda}/n_{e}n_{\sHe^+}$}             &
  \colhead{$4\pi j_{\lambda}/n_{e}n_{\sHe^+}$}             &
  \colhead{}                                               \\
}
\startdata
2945    &    $5\,{}^{3}P$    &    $2\,{}^{3}S$    &    2.657E-26    &   2.70E-26     &      1.6{\%}   \\
3188    &    $4\,{}^{3}P$    &    $2\,{}^{3}S$    &    5.564E-26    &   5.62E-26     &      1.0{\%}   \\
3614    &    $5\,{}^{1}P$    &    $2\,{}^{1}S$    &    6.864E-27    &   6.78E-27     &     -1.2{\%}   \\
3889    &    $3\,{}^{3}P$    &    $2\,{}^{3}S$    &    1.381E-25    &   1.37E-25     &     -0.8{\%}   \\
3965    &    $4\,{}^{1}P$    &    $2\,{}^{1}S$    &    1.398E-26    &   1.39E-26     &     -0.5{\%}   \\
4026    &    $5\,{}^{3}D$    &    $2\,{}^{3}P$    &    2.901E-26    &   2.86E-26     &     -1.4{\%}   \\
4121    &    $5\,{}^{3}S$    &    $2\,{}^{3}P$    &    2.491E-27    &   2.46E-27     &     -1.3{\%}   \\
4388    &    $5\,{}^{1}D$    &    $2\,{}^{1}P$    &    7.677E-27    &   7.58E-27     &     -1.3{\%}   \\
4438    &    $5\,{}^{1}S$    &    $2\,{}^{1}P$    &    1.004E-27    &   1.05E-27     &      4.4{\%}   \\
4472    &    $4\,{}^{3}D$    &    $2\,{}^{3}P$    &    6.110E-26    &   6.16E-26     &      0.8{\%}   \\
4713    &    $4\,{}^{3}S$    &    $2\,{}^{3}P$    &    6.429E-27    &   6.47E-27     &      0.6{\%}   \\
4922    &    $4\,{}^{1}D$    &    $2\,{}^{1}P$    &    1.651E-26    &   1.64E-26     &     -0.7{\%}   \\
5016    &    $3\,{}^{1}P$    &    $2\,{}^{1}S$    &    3.508E-26    &   3.49E-26     &     -0.5{\%}   \\
5048    &    $4\,{}^{1}S$    &    $2\,{}^{1}P$    &    2.417E-27    &   2.53E-27     &      4.5{\%}   \\
5876    &    $3\,{}^{3}D$    &    $2\,{}^{3}P$    &    1.651E-25    &   1.69E-25     &      2.3{\%}   \\
6678    &    $3\,{}^{1}D$    &    $2\,{}^{1}P$    &    4.691E-26    &   4.79E-26     &      2.1{\%}   \\
7065    &    $3\,{}^{3}S$    &    $2\,{}^{3}P$    &    2.867E-26    &   2.96E-26     &      3.1{\%}   \\
7281    &    $3\,{}^{1}S$    &    $2\,{}^{1}P$    &    8.716E-27    &   8.99E-27     &      3.0{\%}   \\
9464    &    $5\,{}^{3}P$    &    $3\,{}^{3}S$    &    1.469E-27    &   1.48E-27     &      0.7{\%}   \\
10830   &    $2\,{}^{3}P$    &    $2\,{}^{3}S$    &    2.720E-25    &   3.40E-25     &     20.0{\%}   \\
11969   &    $5\,{}^{3}D$    &    $3\,{}^{3}P$    &    2.926E-27    &   2.90E-27     &     -0.9{\%}   \\
12527   &    $4\,{}^{3}P$    &    $3\,{}^{3}S$    &    1.782E-27    &   1.79E-27     &      0.5{\%}   \\
12785   &    $5\,{}^{3}F$    &    $3\,{}^{3}D$    &    9.480E-27    &   9.36E-27     &     -1.3{\%}   \\
12790   &    $5\,{}^{1}F$    &    $3\,{}^{1}D$    &    3.158E-27    &   3.14E-27     &     -0.6{\%}   \\
12968   &    $5\,{}^{1}D$    &    $3\,{}^{1}P$    &    9.714E-28    &   9.86E-28     &      1.5{\%}   \\
15084   &    $4\,{}^{1}P$    &    $3\,{}^{1}S$    &    7.429E-28    &   7.39E-28     &     -0.5{\%}   \\
17002   &    $4\,{}^{3}D$    &    $3\,{}^{3}P$    &    4.321E-27    &   4.07E-27     &     -6.2{\%}   \\
18685   &    $4\,{}^{3}F$    &    $3\,{}^{3}D$    &    2.261E-26    &   2.22E-26     &     -1.9{\%}   \\
18697   &    $4\,{}^{1}F$    &    $3\,{}^{1}D$    &    7.535E-27    &   7.39E-27     &     -2.0{\%}   \\
19089   &    $4\,{}^{1}D$    &    $3\,{}^{1}P$    &    1.525E-27    &   1.54E-27     &      0.9{\%}   \\
19543   &    $4\,{}^{3}P$    &    $3\,{}^{3}D$    &    1.039E-27    &   1.05E-27     &      1.0{\%}   \\
21118   &    $4\,{}^{3}S$    &    $3\,{}^{3}P$    &    9.815E-28    &   9.86E-28     &      0.5{\%}   \\
\enddata
\label{table:BSS-Comp}
\end{deluxetable}